\documentclass{aa}  

\usepackage{graphicx}
\usepackage{xcolor}
\usepackage[colorlinks = true, linkcolor = blue, urlcolor  = blue, citecolor = blue]{hyperref}
\usepackage{placeins}
\usepackage{makecell}
\usepackage{txfonts}
\usepackage{booktabs}

\newcommand{\og}{{\tt OpenGADGET3}}
\newcommand{\PINOCCHIO}{{\tt PINOCCHIO}}
\newcommand{\vide}{{\tt VIDE}}
\newcommand{\ZOBOV}{{\tt ZOBOV}}
\newcommand{\PM}{{\tt PM}}
\newcommand{\TreePM}{{\tt TreePM}}
\newcommand{\SUBFIND}{{\tt SUBFIND}}
\newcommand{\FOF}{{\tt FoF}}
\newcommand{\frag}{{\tt FRAGMENTATION}}
\newcommand{\orcid}[1]{}

\definecolor{red1}{RGB}{180,1,25}

\begin{document}

   \title{Tracing cosmic voids with fast simulations}

   \author{M. D. Lepinzan
          \inst{1,2,3,6,7}\orcid{0000-0003-1287-9801}\thanks{\email{marius.lepinzan@inaf.it}},
          C. T. Davies
          \inst{4},
          T. Castro
          \inst{2,3,6,7},
          N. Schuster
          \inst{5},
          J. Mohr
          \inst{4},
          P. Monaco
          \inst{1,2,3,6,7}
          }

   \institute{
   Dipartimento di Fisica, Sezione di Astronomia, Universit\`a di Trieste, Via Tiepolo 11, I-34143 Trieste, Italy \and
   INAF -- Osservatorio Astronomico di Trieste, via Tiepolo 11, I-34131 Trieste, Italy \and
   ICSC -- Centro Nazionale di Ricerca in High Performance Computing, Big Data e Quantum Computing / Spoke 3, Astrophysics and Cosmos Observations, Via Magnanelli 2, Bologna, Italy \and
   University Observatory, LMU Faculty of Physics,  Scheinerstr. 1, 81679 Munich, Germany\and
   Aix-Marseille Universit\'e, CNRS/IN2P3, CPPM, Marseille, France \and
   INFN -- Sezione di Trieste, I-34100 Trieste, Italy \and
   IFPU -- Institute for Fundamental Physics of the Universe, via Beirut 2, 34151, Trieste, Italy
    }
   \date{}

  \abstract
   {Cosmic voids are vast underdense regions in the cosmic web that encode crucial information about structure formation, the composition of the Universe, and its expansion history. Due to their lower density, these regions are less affected by non-linear gravitational dynamics, making them suitable candidates for analysis using semi-analytic methods.}
   {We assess the accuracy of the \PINOCCHIO\ code, a fast tool for generating dark matter halo catalogs based on Lagrangian perturbation theory, in modeling the statistical properties of cosmic voids. We validate this approach by comparing the resulting void statistics measured from \PINOCCHIO\ to those obtained from $N$-body simulations.}
   {We generate a set of simulations using \PINOCCHIO\ and \og, assuming a fiducial cosmology and varying the resolution. For a given resolution, the simulations share the same initial conditions between the two codes. Snapshots are saved at multiple redshifts and post-processed using the watershed void finder \vide\ to identify cosmic voids. For each simulation, we measure the following statistics: void size function, void ellipticity function, core density function, and the void radial density profile. We use these statistics to quantify the accuracy of \PINOCCHIO\ relative to \og\ in the context of cosmic voids.}
   {We find agreement for all void statistics at better than 2$\sigma$ between \PINOCCHIO\ and \og, with no systematic difference in redshift trends. This demonstrates that the \PINOCCHIO\ code can reliably produce void statistics with high computational efficiency compared to full $N$-body simulations.}
   {}

    \titlerunning{Cosmic voids with fast simulations}
    \authorrunning{Lepinzan  et al.}
   \maketitle

\section{Introduction}

Cosmic voids, vast underdense regions in the large-scale structure of the Universe, occupy the majority of its volume. They are promising cosmological probes that offer complementary and improved constraints on cosmological models~\citep{Pisani:2019cvo, Davies2021, Moresco:2022phi}, and their capacity to inform cosmological constraints will continue to grow in this era of large-scale surveys~\citep{Euclid:2021xmh, Contarini:2022mtu, Euclid:2023eom}. Their underdense nature not only makes them less affected by non-linear gravitational dynamics~\citep{Hamaus:2014fma,Schuster:2022ogh}, providing a unique window into the growth of structures and the properties of the Universe's expansion~\citep{Hamaus:2016wka,Schuster:2025}, but also provides unique sensitivity to the underlying cosmological model.

From a theoretical perspective, the evolution of cosmic voids can be effectively described using the excursion-set formalism, which provides an analytical framework for modeling their formation and statistical properties. Unlike overdense structures, voids expand quasi-linearly, making them particularly well-suited for this approach. Their evolution follows two key processes: the void-in-void mechanism, where smaller voids merge into larger ones, and the void-in-cloud process, where voids embedded in initially overdense regions are compressed and eventually erased~\citep{Sheth:2003py, Jennings:2013nsa,Schuster:2025}.

In addition to providing insights into structure formation, voids serve as powerful probes for testing fundamental physics. Their shapes evolve under the combined influence of tidal distortions and gravitational dilution, making them powerful probes for constraining the dark energy equation of state~\citep{Lee:2007kq, Bos:2012wq}. Furthermore, the size abundance of voids also depends on the dark energy equation of state: as matter flows out of voids, they expand while surrounding structures collapse, making their size and evolution inherently sensitive to the effects of cosmic acceleration~\citep{Pisani:2015jha}.

Moreover, voids offer an ideal setting for testing modified gravity theories \citep{Li2011, Davies2019}. In these models, deviations from general relativity are suppressed by screening mechanisms which are generally present in overdense environments, but these same screening mechanisms becomes ineffective in underdense enviroments. This makes voids particularly effective in detecting deviations from General relativity (GR)~\citep{Perico:2019obq}. Additionally, void statistics can help disentangle the degeneracy between modified gravity and the neutrino mass sum, probing their combined effects on cosmic structure formation~\citep{Contarini:2020fdu}.

Using precise and computationally efficient tools is crucial for studying the large-scale structure of the universe and conducting the tests outlined above. Fast semi-analytic approaches, such as \PINOCCHIO\ \citep{Monaco:2001jg, Monaco:2013qta}, leverage Lagrangian perturbation theory (LPT) to rapidly generate halo distributions without the computational overhead of full $N$-body simulations. Previous studies have demonstrated the accuracy of \PINOCCHIO\ in reproducing halo statistics, such as the halo mass function (HMF), halo bias, and 2-point statistics~\citep[see, for instance,][]{Monaco:2013qta,Munari:2016aut, Fumagalli:2025twg, Euclid:2025lfa}. However, its effectiveness in modeling other components of the cosmic web remains an open question. 

To test the reliability of \PINOCCHIO\ in predicting void statistics, we use the full $N$-body code \og\ as a benchmark. In this work, voids are identified using the \vide\ toolkit, which is applied to the halo field (rather than the full matter density field) constructed from the halo catalogs of both simulations. By comparing results across different redshifts and resolutions, we assess how well \PINOCCHIO\ reproduces the statistical and structural properties of voids over cosmic time. We investigate four key summary statistics that characterize cosmic voids: the void size function (VSF), void ellipticity function (VEF), core density function (CDF), and radial density profiles (RDP). The investigation into each of these statistics is motivated below.

As a complementary tool to the analysis of overdense structures such as halos, voids provide an alternative perspective on structure formation. While halos grow through mass accretion and gravitational collapse, reducing their comoving volume, voids expand. Consequently, the variation of the VSF as a function of redshift captures the evolution of the Large-scale structure (LSS) from a unique perspective. By tracing the size distribution of these expansive voids, the VSF contains valuable information on various cosmological parameters, complementing standard probes~\citep{Pisani:2015jha, Contarini:2019qwf, Contarini:2022mtu}, which typically focus on overdensities.

Beyond providing insights into individual void morphologies, the resulting VEF carries crucial cosmological information. The anisotropic growth of cosmic structures, shaped by tidal forces and surrounding matter distributions~\citep{Park:2006wu, Schuster:2022ogh}, directly influences void deformation. As matter collapses into cosmic structures the surrounding voids deform accordingly. Since void measurements are largely unaffected by systematics from baryonic physics~\citep{Schuster:2024}, the redshift evolution of their shape distribution serves as a valuable tracer of dark energy~\citep{Lee:2007kq,Bos:2012wq,Schuster:2025}. Additionally, the average stretching of voids along the line of sight enables tests of cosmic expansion using the Alcock-Paczynski effect~\citep{Alcock:1979mp, Sutter:2014oca}.

The CDF encodes information about the variations in void emptiness, the processes driving matter evacuation, and the interaction of voids with their surrounding structures. Therefore the CDF provides valuable insights into the underlying cosmology, particularly in scenarios involving massive neutrinos \citep{Schuster:2019hyl, Schuster:2022ogh}.

The stacked RDP is powerful probe of fundamental physics, particularly in testing deviations from GR and probing the late-time evolution of the Universe. Screening mechanisms suppress the fifth force in overdense regions, but become inefficient in void interiors, and deviations from GR therefore manifest in the RDP. In particular, modified gravity models often predict enhanced void expansion, leading to deeper void centers and steeper compensation walls compared to the predictions of $\Lambda$CDM~\citep{Perico:2019obq, Contarini:2020fdu}. The shape of the RDP is also directly linked to the Integrated Sachs-Wolfe (ISW) effect~\citep{Sachs:1967er}, as voids dynamically evolve within the cosmic web. In $\Lambda$CDM, decaying void potentials cause a colder ISW imprint, modifying the expected signal~\citep{Ilic:2013cn, Kovacs:2018}. 

This work is structured as follows. We begin with a description of the void finder used in this work (Sect.~\ref{VOID_Finder}), followed by an overview of the main void statistics that we investigate (Sect.~\ref{VOID_statistics}). The simulation setup is outlined in Sect. \ref{Cosmological_Simulation}, where we also describe the two codes employed in this study, \og\ (Sect. ~\ref{Gadget}) and \PINOCCHIO\ (Sect.~\ref{PINOCCHIO}). A comparison of the HMFs derived from both codes is then presented in Sect.~\ref{HMF_comparison_section}, including details on the mass correction applied to \og\ halos and the procedure for matching halo number densities. In Sect.~\ref{VOID_Results}, we present a detailed comparison of the void statistics between the two simulations. Finally, the main conclusions of our study are summarized in Sect.~\ref{Conclusions}

\section{Methodology}

\subsection{Void identification}
\label{VOID_Finder}
In this work, we employ the Void Identification and Examination toolkit \vide~\footnote{\url{https://bitbucket.org/cosmicvoids/vide_public/}}~\citep{Sutter:2014haa} to identify cosmic voids from a set of biased tracers (halos) for our subsequent analysis. \vide\ implements an enhanced version of the ZOnes Bordering On Voidness (\ZOBOV) algorithm~\citep{Neyrinck:2007gy}, which is based on a watershed technique~\citep{Platen:2007qk} that identifies local watershed basins in a given $3D$ density field, and is outlined as follows. The process begins with a Voronoi tessellation of the tracer positions, which is then used to estimate the density field. The density at each point is calculated as the inverse of the volume of its corresponding Voronoi cell. Each cell is then grouped with its corresponding neighbor cell with the lowest density value of all neighbors. This step is repeated until all cells are assigned to a group, where each group contains one local minima. These groups of cells are the watershed basins, which correspond to the identified void population.

Once the watershed step is complete, each void is represented by a collection of Voronoi cells with an arbitrarily irregular overall shape. Each void's center is determined by the volume-weighted barycenter of all the Voronoi cells associated with the void. This is calculated by summing over the comoving positions $x_j$ of the tracers, weighted by the volumes of their associated cells:
\begin{equation}
\label{void_center}
X_V = \frac{\sum_j x_j V_j}{\sum_j V_j} \,.
\end{equation}
The total volume for each void is calculated as the sum of the volumes of the Voronoi cells $j$ that belong to that void. The effective radius $R_{\mathrm{eff}}$ is then defined as the radius of a sphere with an equivalent volume $V$: 
\begin{equation}
\label{effective_radius}
R_{\mathrm{eff}} = \left(\frac{3}{4\pi}\sum_j V_j\right)^{1/3} \,.
\end{equation}

Additionally, \vide\ quantifies the shape of the voids by computing their inertia tensor,
\begin{equation}
\label{inertia_tensor}
\begin{aligned}
M_{xx} &= \sum_j\left(y^2_j + z^2_j\right) \,, \\
M_{xy} &= - \sum_j\left(x_j\,y_j\right) \,,
\end{aligned}
\end{equation}
with $x_j$, $y_j$, and $z_j$ representing the co-moving coordinates of the tracers relative to the void center defined in Eq. \eqref{void_center}. The remaining components of the inertia tensor are calculated similarly to Eq. \eqref{inertia_tensor}. 

\noindent From this, the void ellipticity is then defined in terms of the smallest $J_1$ and largest $J_3$ eigenvalues of the inertia tensor:
\begin{equation}
\label{void_ellipticity}
\epsilon = 1 - \left(\frac{J_1}{J_3}\right)^{1/4} \,.
\end{equation}

Another key void property, is the core density, denoted as $\hat{n}_{C}$,
\begin{equation}
\label{core_density}
\hat{n}_{C} = \frac{n_{\text{core}}}{\Bar{n}_t} \,,
\end{equation} 

\noindent which represents the density of the largest Voronoi cell of a void, $n_{\text{core}}$, corresponding to the region of lowest density of a particular void. This quantity, computed by \vide\, is expressed relative to the mean density of the tracer population $\Bar{n}_t$.

Lastly, to calculate void radial tracer density profiles, following~\citep{Hamaus:2014fma, Schuster:2022ogh}, the (number) density within radial shells of thickness $2\delta r$ at a given co-moving distance $r$ from the center of a single void \textit{i} can be defined as,
\begin{equation}
\label{density_in_shell}
{\rho}_{V}^{(i)}(r) = \frac{3}{4\pi} \sum_j \frac{\Theta(r_j)}{(r + \delta r)^3 - (r - \delta r)^3} \,,
\end{equation} 

\noindent where $\Theta (r_j)$ is defined through two Heaviside step functions $\vartheta$, which specify the radial bin:
\begin{equation}
\Theta (r_j) \equiv \vartheta [r_j - (r - \delta r)]\vartheta [-(r_j - \delta r)^3] \,.
\end{equation}

\noindent Here, $r_j$ denotes the co-moving distance of the \textit{j}-th tracer from the void center, while $\delta r$ determines the shell thickness. The summation in Eq. \eqref{density_in_shell} includes all tracers \textit{j} within a specified distance from the void. 

\subsection{Void statistics}
\label{VOID_statistics}
This subsection introduces the key void summary statistics derived from the \vide\ output: VSF, VEF, CDF and RDP. In all cases we analyze these properties at redshifts $z \in \{0.0 , 0.5 , 1.0 , 1.5, 2.0\}$.
\paragraph{VSF:} similar to the HMF, which describes the number density of collapsed objects as a function of their mass and redshift~\citep{Press:1973iz, Sheth:1999mn,Euclid:2022dbc}, the VSF describes the number density of cosmic voids as a function of their size $R_{\mathrm{eff}}$ Eq. \eqref{effective_radius} and redshift~\citep{Sheth:2003py,Jennings:2013nsa}. We measure the VSF with 17 linearly spaced bins between 10 $h^{-1}\,$Mpc and 80 $h^{-1}\,$Mpc.
  
\paragraph{VEF:} although characterizing the nonspherical shape of an unbound system like a void is challenging due to the lack of a clearly defined boundary, the VEF serves as a good first-order probe of deviations from spherical symmetry~\citep{Park:2006wu}. Void shapes are commonly characterized using the inertia tensor (Eq.~\eqref{inertia_tensor}). The departure from sphericity can be quantified by measuring the ellipticity (Eq.~\eqref{void_ellipticity}) of the spatial distribution of the void tracers. We measure the VEF with 11 linearly spaced bins between $\epsilon = 0$ and $\epsilon = 0.4$, as voids at higher ellipticity are extremely sparse.

\paragraph{CDF:} similar to the HMF and VSF, the CDF describes the number density of cosmic voids as a function of their minimal density $n_C$ (Eq.~\ref{core_density}), offering insights into the extreme under-dense environments that define their cores~\citep{Schuster:2024}. We measure the CDF with 12 linearly spaced bins between $n_C = 0.05$ and $n_C = 0.65$.
  
\paragraph{RDP:} the RDP of a single void measures the density contrast relative to the mean tracer density, which is defined as,
\begin{equation}
RDP_i(r) = {\rho}_{V}^{(i)}(r)/\Bar{\rho} - 1 \,,
\end{equation}
where ${\rho}_{V}^{(i)}(r)$ is the void density within a radial shell, as described in Eq.~\eqref{density_in_shell}, and $\Bar{\rho}$ is the mean tracer density. We emphasize that, in this work, the density field is inferred using the number density of halos, rather than the underlying full dark matter distribution. This means that the RDP characterizes the distribution of halos (tracers) around voids rather than the total mass content.

Whether measured from halos or the full density field, the RDPs of individual voids are scattered and affected by the underlying resolution. While they do not individually contain much cosmological information, they remain useful for testing specific void characteristics~\citep{Schuster:2022ogh, Schuster:2024}. Furthermore, a more robust and representative characterization of the void profiles can be obtained by stacking the individual void profiles. 
   
The stacked profiles are obtained by first computing the radial density profile for each void individually, then grouping the voids into bins based on their effective radius $R_{\mathrm{eff}}$ (Eq.~\ref{effective_radius}). The profiles within each bin are averaged to produce a representative stacked profile for that void size range. Given this approach, the final stack is an average of Eq. \eqref{density_in_shell},
\begin{equation}
\rho_{V}(r) = \frac{1}{N_v} \sum_i \rho_V^{(i)}(r) \,,
\end{equation}
accordingly, the resulting RDP is redefined as:
\begin{equation} RDP(r) = \rho_V(r)/\Bar{\rho} - 1 . 
\end{equation}
This approach allows us to dissect trends in the void halo distribution over the whole void population and over a range of scales, reducing the overall scatter from individual voids~\citep{Hamaus:2014fma, Schuster:2022ogh}. We measured the individual RDPs for each void out to $2.5 \times R_{\mathrm{eff}}$ using 12 linearly spaced bins, and grouped them into 3 linearly spaced bins of void size reported in Table~\ref{tab:void_bins}.

\subsection{Cosmological simulations}
\label{Cosmological_Simulation}
The cosmological simulations used in this study are generated from two distinct methods and codes, an $N$-body code and a perturbation theory based code. The first is the $N$-body code \og\ (Dolag et al in prep.) described in Sect.~\ref{Gadget}. 
The latter is the LPT code \PINOCCHIO~\footnote{\url{https://github.com/pigimonaco/Pinocchio}}~\citep{Monaco:2013qta} described in Sect.~\ref{PINOCCHIO}.

For both simulations we use a fixed box size of 512 $h^{-1}\,$Mpc and two particle resolutions: $512^3$ and $1024^3$. To ensure consistency in comparing the two approaches, the initial conditions (ICs) used by both simulations were generated with \PINOCCHIO\ at redshift $z = 50$ using third-order Lagrangian perturbation theory (3LPT). This allows us to trace the evolution of the same initial density fields across redshift with different simulation codes.

The simulation outputs were recorded at five different redshifts  $z = (0.0 , 0.5 , 1.0 , 1.5, 2.0)$, with the goal of evaluating the accuracy with which PINOCCHIO can replicate the statistical properties of cosmic voids in a full $N$-body simulation, at various stages in cosmic evolution. 

Both simulations assume a flat $\Lambda$CDM cosmology that matches Planck15~\citep{Planck:2015fie}. The matter density parameter is set to $\Omega_{\rm m} = 0.315$, which gives a dark energy density of $\Omega_{\Lambda} = 0.685$. The cosmic baryon density is $ \Omega_{\rm b} = 0.022\,h^{-2}$, with the dimensionless Hubble parameter $h = 0.673$. Finally, the linear power‑spectrum is normalized at redshift $z = 0$ by $\sigma_{8} = 0.829$, and the primordial spectral index is $n_{\rm s} = 0.966$.

\subsubsection{OpenGADGET3}
\label{Gadget}
\og\ (Dolag et al in prep.) is a highly flexible and efficient code used for simulating the gravitational and hydrodynamic evolution of cosmic structures, ranging from galaxies to the large-scale structure of the Universe. The core of \og's gravitational dynamics for a collision-less fluid relies on two complementary schemes: a hierarchical tree algorithm~\citep{Barnes:1986nb} for short-range interactions and a Particle-mesh (\PM) method~\citep{Efstathiou:1985re} for long-range contributions. This combination results in a \TreePM~\citep{Xu:1994fk} approach, where the \PM\ method significantly reduces the computational complexity of long-range gravitational interactions from \textit{O}(\textit{N}$^{2}$) to \textit{O}(\textit{N}log\textit{N}), while the tree algorithm efficiently resolves local gravitational dynamics.  
These optimizations allow \og\ to accurately and efficiently simulate large cosmological volumes~\citep{Springel:2005mi}. Although \og\ includes Smoothed particle hydrodynamics (SPH) for modeling complex gas processes, this work focuses exclusively on DM dynamics, because the void scales investigated in this work are largely unaffected by these more complex processes~\citep{Schuster:2024, Lehman:2025}. 

In this work, the ICs are generated using \PINOCCHIO\ and provided as input to \og. The particle distribution then evolves under the influence of gravity, as described above. The tree structure is updated periodically to track the movements of the particles throughout the simulation volume. At each time step, particle positions and velocities are updated based on the gravitational forces using a leapfrog integration scheme~\citep{Duncan_1998}.

The final halo catalog in \og\ is generated on the fly using the \SUBFIND\ algorithm~\citep{Springel:2000qu, Dolag:2008ar}. \SUBFIND\ operates in two steps: first, it uses a Friends-of-Friends (\FOF) algorithm to identify parent haloes. Second, \SUBFIND\ detects subhalos by estimating densities via SPH, locating overdense regions, and applying a gravitational unbinding procedure to retain only self-bound structures.

\subsubsection{PINOCCHIO}
\label{PINOCCHIO}

\PINOCCHIO~\citep[PINpointing Orbit Crossing-Collapsed HIerarchical Objects;][]{Monaco:2001jg, Monaco:2013qta, Munari:2016aut, Euclid:2025lfa} is a code designed to produce fast simulations of the distribution of dark matter halos by employing approximate methods, starting from an initial density field. \PINOCCHIO\ combines the Extended Press-Schechter approach~\citep{Press:1973iz,Bond:1990iw}, and its generalization to a more realistic ellipsoidal collapse scenario~\citep{Monaco:1997cq}. 

The calculation begins by generating a linear density field as a realization of a Gaussian random process; this is then smoothed over a series of decreasing radii $R$. The collapse time for each grid point is then computed using ellipsoidal evolution~\citep{Monaco:1997cq}, approximated by 3LPT. It is well known that the LPT dynamics are accurate up to orbit crossing events, making it suitable for predicting the moment when the ellipsoid collapses along its shortest axis~\citep{Monaco:1994ed}. At this stage, the particle (or equivalently the grid point) is expected to become part of a DM halo or of the filamentary network connecting halos~\citep{Shandarin:1989sr}.

In order to group collapsed particles into halos or filaments and construct halo merger histories, a custom algorithm called \frag\ is applied~\citep{Monaco:2013qta}. This algorithm mimics the hierarchical process of matter accretion and halo merging. The accretion of particles onto a halo, as well as the merging of two halos, is determined on the basis of a specific criterion. The object pair (halo-particle or halo-halo) is mapped from their initial Lagrangian positions to their predicted Eulerian positions at a given time using LPT displacements. Accretion or merging occurs when their separation $d$ falls below a given threshold $d{\rm_{thr}}$, which depends on the Lagrangian radius of the larger object. 

LPT displacements are applied in two key steps: first, during halo construction, and second, when determining their final Eulerian positions for the output catalog~\footnote{The displacement of a halo is computed as the average displacement of all particles that belong to it.}. To achieve the best agreement with full $N$-body simulations, it has been shown~\citep{Munari:2016aut} that using 3LPT for halo displacements and 2LPT for halo construction yields results within 10\% accuracy in the halo power spectrum (up to $k_{\rm max} \sim 0.5 h\,{\rm Mpc}^{-1}$) and within 5\% - 10\% in the HMF (up to $\sim10^{15}M_\odot$). 

Due to its approximate yet robust approach and the highly parallelized structure of most of its components~\citep{Monaco:2013qta}, \PINOCCHIO\ is capable of generating realistic cosmic structure simulations at a small fraction of the computational cost compared to full $N$-body simulations. With a good degree of accuracy for the halo mass function and a computational cost more than 1.000 times lower~\citep{Monaco:2013qta,Munari:2016aut}, it offers a highly efficient alternative for generating many realizations of the Universe, making it especially suitable for studies requiring numerous realizations, large volumes, or rapid parameter exploration.

\subsubsection{Halo mass function: OpenGADGET3 vs PINOCCHIO}
\label{HMF_comparison_section}

Before analyzing voids, it is essential to first examine the halo populations, as the halos serve as tracers for void identification. Ensuring consistency in the halo catalogs between the two simulations is therefore crucial for a fair comparison of the resulting void statistics.

First, we perform some preliminary cleaning of the halo catalogs as output by the two simulation codes. This involves addressing discrepancies in the halo definitions between \og\ and \PINOCCHIO. This cleaning process consists of two main steps.
\begin{figure*}[t]
\centering
\includegraphics[width=0.97\textwidth]{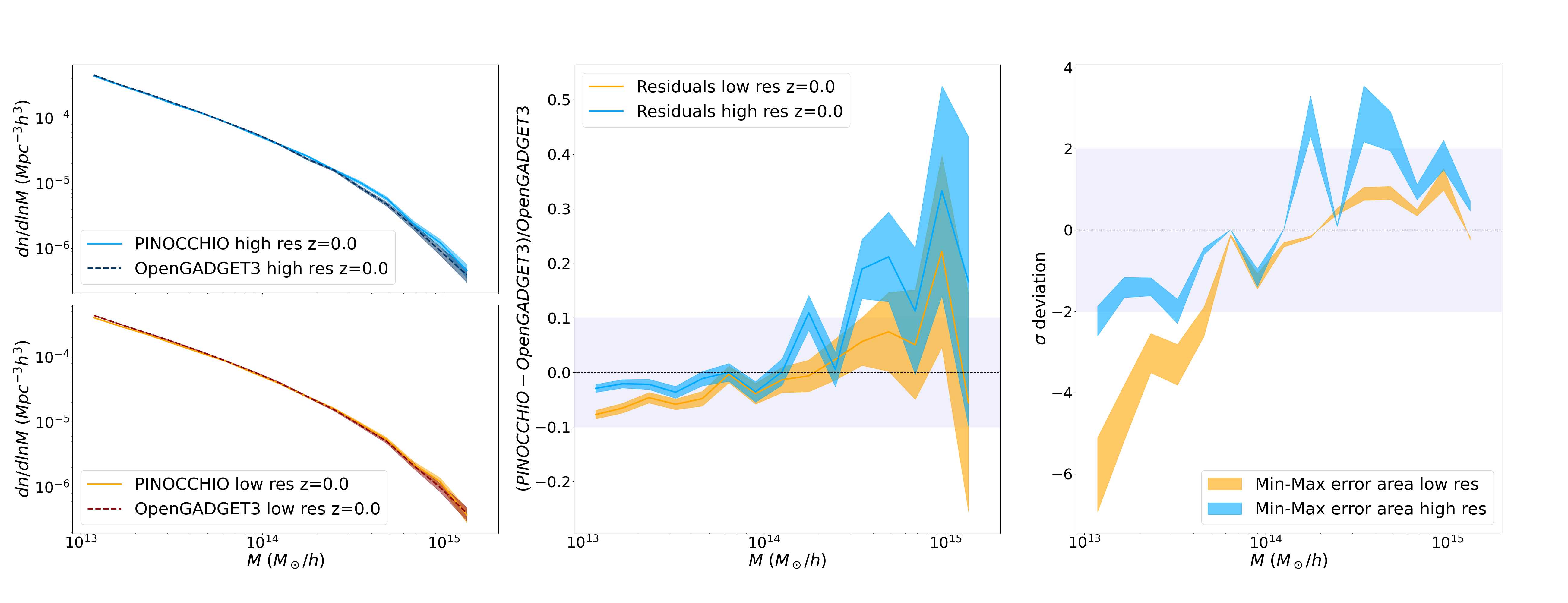}
\caption{Comparison of the HMFs (left panel) at $z=0.0$ for low- and high-resolution simulations. Jackknife errors are shown as filled regions. In the middle panel, the shaded band indicates a $\pm 10\%$ range around the \og\ result, offering a reference for the level of agreement with \PINOCCHIO. The filled regions indicates the Jackknife errors around the residual curves. The right panel displays the relative difference between the two HMFs, normalized by the maximum and minimum statistical uncertainties, as described in Eqs. \eqref{eq:errors_min} and \eqref{eq:errors_max}, and expressed in units of $\sigma$. The shaded region indicates the $\pm2\sigma$ range, highlighting the level of statistical consistency between the two HMFs. The filled region further illustrates the variation range between these two error estimates.}
\label{HMF_comparison}
\end{figure*}
\paragraph{FoF Mass Consistency:} Since \PINOCCHIO\ has been calibrated to match the Watson model~\citep{Watson:2012mt}, which is based on the \FOF\ halo mass definition, we ensure consistency by also using \FOF\ masses for \og\ halos. However, the \FOF\ algorithm is known for overestimating halo masses, particularly for halos with a low particle count. To mitigate this bias, we apply the correction proposed by~\citep{Warren:2005ey}, which adjusts the particle count as follow,
\begin{equation}
N_{\rm corrected} = N(1 - N^{-0.6}) \,,
\end{equation}
where $N$ is the number of particles in the halo. This correction is applied to the \og\ halo masses in our analysis.

\paragraph{Number Density Matching:} Although \PINOCCHIO\ is calibrated to match the \FOF\ HMF, its accuracy is limited by systematic differences between the two codes, particularly at the low-mass end, and motivates the need for further calibration.

Given that void statistics depend on the properties of the underlying tracer distribution used to define the voids, to ensure a fair comparison between \og\ and \PINOCCHIO\, we also match the number density of halos in the two halo catalogues. 
    
To achieve this, we first apply a mass threshold cut of $10^{13}M_\odot/h$ to the corrected \og\ masses, and count the number of halos above this threshold. We then select the same number of halos from the \PINOCCHIO\ catalog, sorted in decreasing mass, to construct the corresponding \PINOCCHIO\ catalog. This number density matching serves as a halo catalog calibration, analogous to the procedure used in \cite{Euclid:2021api}, because both simulations in our analysis share the same ICs.

This calibration is necessary because the HMF is sensitive to differences in the methods used to generate halo catalogs. These include differences in simulation codes, halo definitions, and numerical resolution~\citep{Euclid:2022dbc}. Therefore, calibrating the HMF by accounting for these systematic differences between codes is crucial for fair comparisons and the robustness of our results. 

The halo mass cut of $10^{13}M_\odot/h$ provides a “golden sample” of voids linked to robustly identified halos, well suited for studies involving luminous red galaxies (LRGs). Extending the framework to lower-mass cut relevant for emission line galaxies (ELGs) will require additional work due to more uncertain halo identification, stronger scale-dependent galaxy-bias, and enhanced non-linear effects. Nonetheless, the stability of \PINOCCHIO\ at higher mass resolution has been tested for other probes, with galaxy clustering accurately reproduced down to $\sim 1.5\cdot 10^{11}M_\odot/h$ \citep{Euclid:2025lfa}, so we do not expect major changes in the void statistics when moving to lower mass cut.

These steps ensure that the halo catalogs from the two simulations are statistically comparable, minimizing the impact of halo number density differences on the resulting void statistics. Because the void finder relies on the spatial distribution of tracers and not their masses, matching the number density, rather than applying identical mass thresholds, provides a more meaningful basis for comparison.

In Figure~\ref{HMF_comparison}, we show the HMFs measured from \og\ and \PINOCCHIO\ after the above corrections and criteria have been applied. The figure includes three panels. The left panel displays the HMFs for both low- and high-resolution simulations ($N_p = 512^3$ and $N_p = 1024^3$, respectively), with their respective Jackknife errors shown as shaded regions. \og\ results are shown using dashed lines, while \PINOCCHIO\ is represented by solid lines; different colors are used to distinguish the two resolutions. The middle column shows the difference between the HMFs, normalized by the \og\ values. The right column presents the difference between the HMFs, normalized by the statistical uncertainties using two estimators,
\begin{equation}
\label{eq:errors_max}
\rm Max_{\text{err}} = \frac{HMF_{\text{\PINOCCHIO}} - HMF_{\text{\og}}}{\sigma_{\text{\og}}} \,,
\end{equation}
\begin{equation}
\label{eq:errors_min}
\rm Min_{\text{err}} = \frac{HMF_{\text{\PINOCCHIO}} - HMF_{\text{\og}}}{\sqrt{\sigma_{\text{\PINOCCHIO}}^2 + \sigma_{\text{\og}}^2}} \,,
\end{equation}
where $\sigma_{\text{\PINOCCHIO}}$ and $\sigma_{\text{\og}}$ are the respective Jackknife errors. Since the two simulations share the same ICs, the statistical uncertainties are positively correlated. Ideally, the significance of their differences should be estimated by accounting for this covariance. In the absence of a direct estimate of the covariance, we adopt two approximations that capture the range of plausible significance values. The $\rm Max_{\text{err}}$ estimator accounts only for the statistical uncertainty of the \og\ , effectively assuming that the \PINOCCHIO\ prediction is exact. This results in an upper-limit of the significance. In contrast, the $\rm Min_{\text{err}}$ estimator assumes uncorrelated errors, thereby overestimating the total variance and yielding a lower-limit significance. Together, these two estimators provide a conservative estimate of the range within which the true significance is expected to lie.

The comparisons in Figure~\ref{HMF_comparison} show the level of agreement between the \og\ and \PINOCCHIO\ HMFs at $z = 0.0$. For both low- and high-resolution simulations, the middle panel demonstrates that \PINOCCHIO\ remains well within the $\pm 10\%$ range of \og\ for most mass scales, consistent with findings in~\cite{Munari:2016aut}. 

The right column shows that most values fall within $\pm2\sigma$, indicating overall consistency with statistical expectations. However, the clear mass-dependent deviations, reaching more than $-3\sigma$ at the low-mass end, point to potential systematic discrepancies. These discrepancies are likely driven by limitations in the number density matching algorithm, which assumes a one-to-one correspondence between structures in \PINOCCHIO\ and \og. This assumption becomes increasingly prone to systematic effects at lower masses, where the identification of halos is inherently noisier. Nonetheless, the agreement between the two methods remains within 10$\%$, suggesting that we are limited more by systematics than by statistical uncertainties in this regime.
\begin{figure}[t]
   \centering
   \includegraphics[width=0.95\columnwidth]{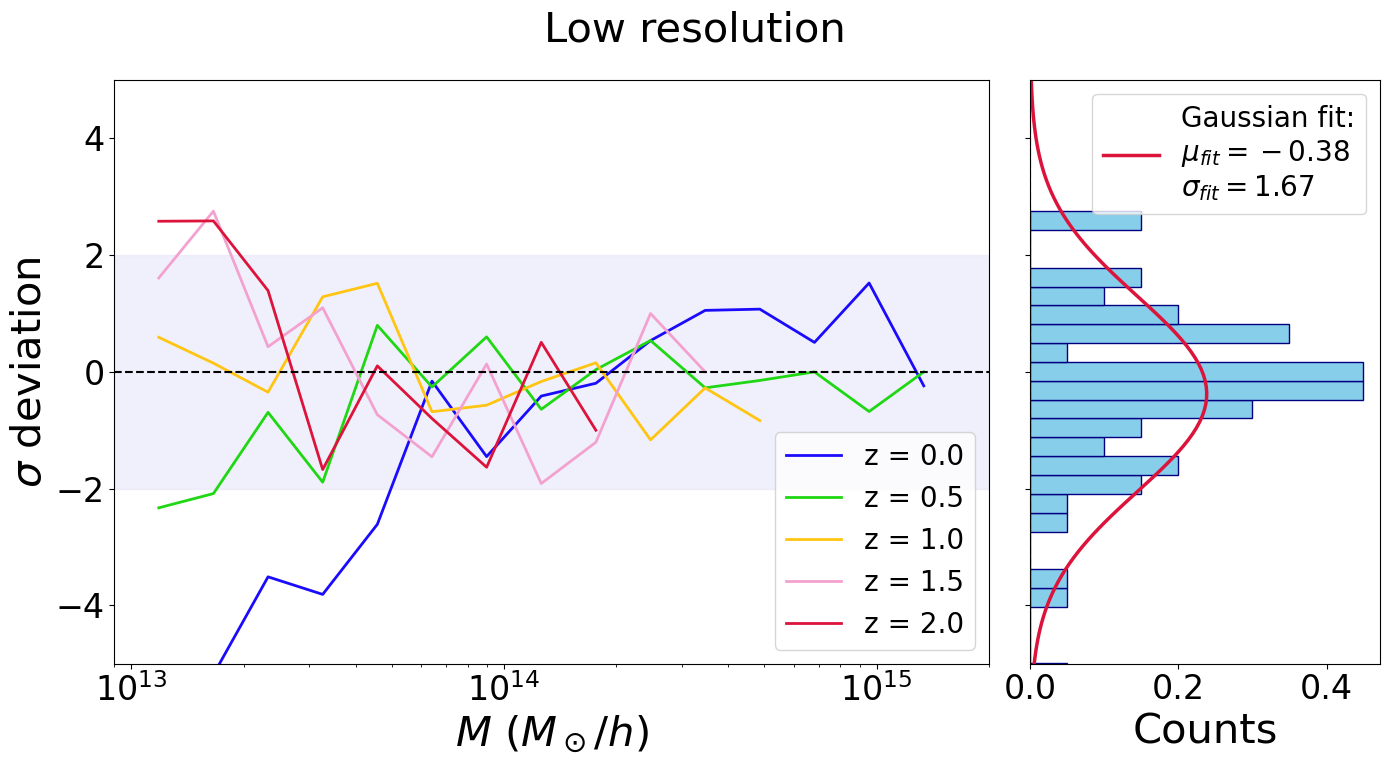}

   \vspace{0.1em}
   
   \includegraphics[width=0.95\columnwidth]{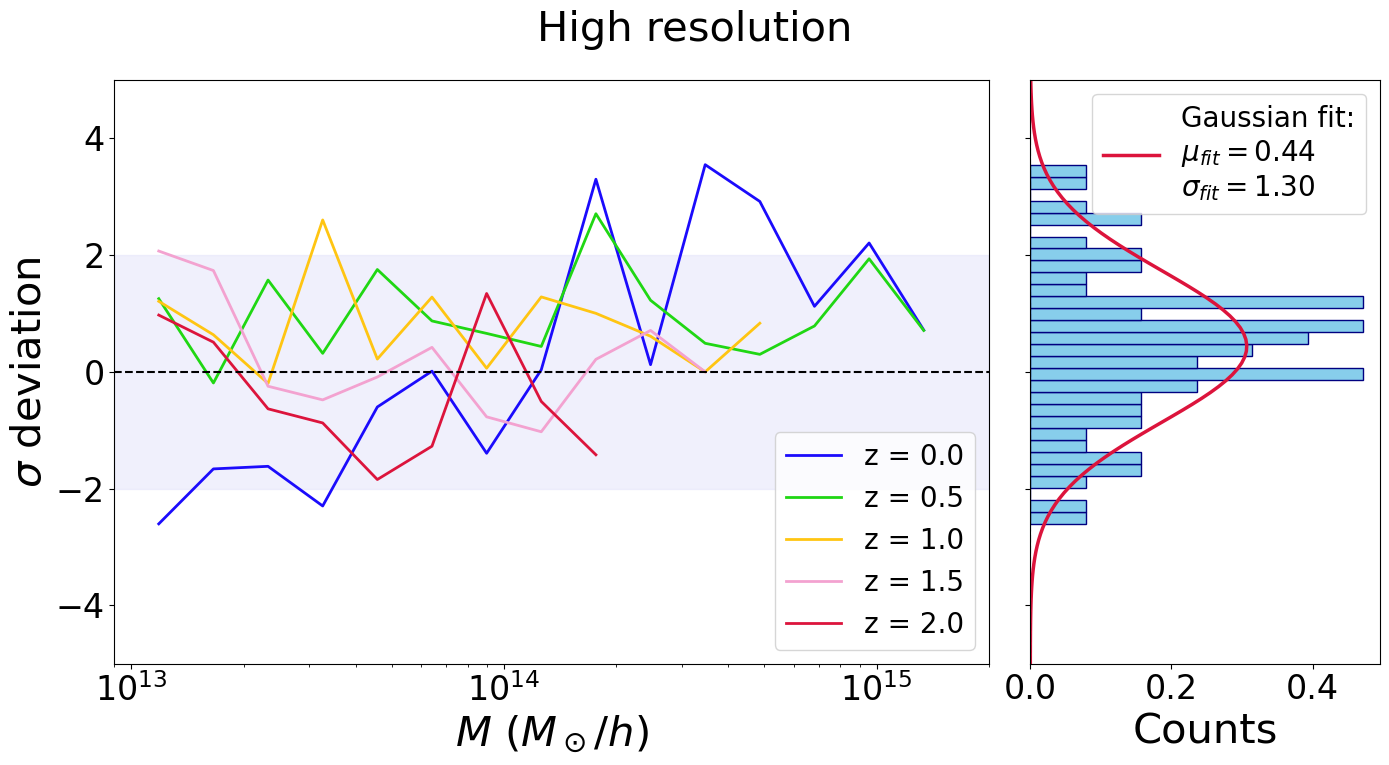}
      
   \caption{Left panels: relative difference between the two HMFs at different redshifts  $z = (0.0 , 0.5 , 1.0 , 1.5, 2.0)$, with low-resolution results shown on the top and high-resolution results on the bottom. Each line represents the difference between \PINOCCHIO\ and \og\, normalized by the Jackknife errors from the \og\ estimates as defined in Eq.~\eqref{eq:errors_max}. The shaded area indicates the $\pm 2\sigma$ range, highlighting the region of statistical agreement between the two methods. The horizontal dashed line at $\sigma = 0$ serves as a reference for perfect agreement. Right panels: overall distribution of the measurements in the left panels, aggregated over all redshifts and size bins, with overlaid Gaussian fits. }
   \label{HMF_evolution}
\end{figure}

Figure~\ref{HMF_evolution} repeats the measurements defined in Eq.~\eqref{eq:errors_max} for multiple redshifts. The figure consists of two panels: the top panel corresponds to measurements from the low-resolution simulations ($N_p = 512^3$), while the bottom panel shows the results from the high-resolution case ($N_p = 1024^3$).
The left panels results confirm that the agreement between \PINOCCHIO\ and \og\ is consistent across a broad range of redshifts and halo masses. The $\pm 2\sigma$ shaded region illustrates that the two methods remain statistically consistent across most mass scales and redshift intervals. However, particularly in the upper panel, deviations reaching up to $\pm3\sigma$ in low mass bins, are observed, similar to the trends seen in the right column of Figure~\ref{HMF_comparison}.
These larger deviations arise not from large absolute discrepancies, but from the small statistical uncertainties at these mass scales. As a result, the $\sigma_{\text{\og}}$-normalized relative difference (Eq.~\eqref{eq:errors_max}) become more sensitive to even modest mismatches. This regime is therefore limited more by systematic effects than by statistical noise.

The right panels provide a complementary, global diagnostic through the distribution of all the deviation values aggregated across mass bins and redshifts. In an ideal case, perfect agreement between the methods would yield a Gaussian with mean $\mu_{fit} = 0$ and width $\sigma_{fit} = 1$, indicating that the residual scatter is entirely due to statistical uncertainties. In both resolution cases, the fitted distributions show $|\mu_{fit}| - \sigma_{fit} < 0$, implying that the residuals do not present a significant bias. Widths broader than unity ($\sigma_{\mathrm{fit}} > 1$), suggest that the observed scatter slightly exceeds what is expected from purely statistical fluctuations. This excess variance likely reflects localized systematics, such as those observed in the low-mass bins, rather than a global mismatch. This supports the conclusion that, despite some small deviations, the overall statistical consistency between \PINOCCHIO\ and \og\ remains robust.

As voids are underdense regions, they are less affected by non-linear gravitational effects, and they are particularly well-suited for semi-analytic methods. Given this context, the same level of agreement observed here is expected for the subsequent void statistics, which we investigate in the next section.

\section{Results}
\label{VOID_Results}

The following sections present summary statistics of cosmic voids, identified with \vide\ (Section \ref{VOID_Finder}) and applied to the halo catalogs of \og\ and \PINOCCHIO. The analysis presented here follows the cleaning procedures described in Section \ref{HMF_comparison_section}.

\subsection{Void size function}
\label{VSF_Section}

The number of voids identified for each redshift $z$ is reported in Table~\ref{tab:void_counts}, showing comparable counts between the two codes in both simulation resolutions. Since the same mass cut is applied at both resolutions, the number of halos remains comparable, leading \vide\ to identify a similar number of voids in both cases.
\begin{table}[t]
\caption{Number of voids identified by \vide}
\centering
\label{tab:void_counts}
\resizebox{0.78\columnwidth}{!}{%
\begin{tabular}{c|cc|cc}
\toprule
$z$ & \multicolumn{2}{c|}{\PINOCCHIO} & \multicolumn{2}{c}{\og} \\
            & Low Res & High Res & Low Res & High Res \\
\midrule
0.0 & 597 & 602 & 583 & 591 \\
0.5 & 468 & 480 & 486 & 516 \\
1.0 & 344 & 340 & 326 & 361 \\
1.5 & 211 & 205 & 209 & 216 \\
2.0 & 107 & 108 & 106 & 113 \\
\bottomrule
\end{tabular}%
}
\tablefoot{The first column lists the redshifts $z$, while the second and third columns shows the number of voids identified for both simulation resolutions and codes.}
\end{table}
\begin{figure*}[t]
    \centering
    \includegraphics[width=0.97\textwidth]{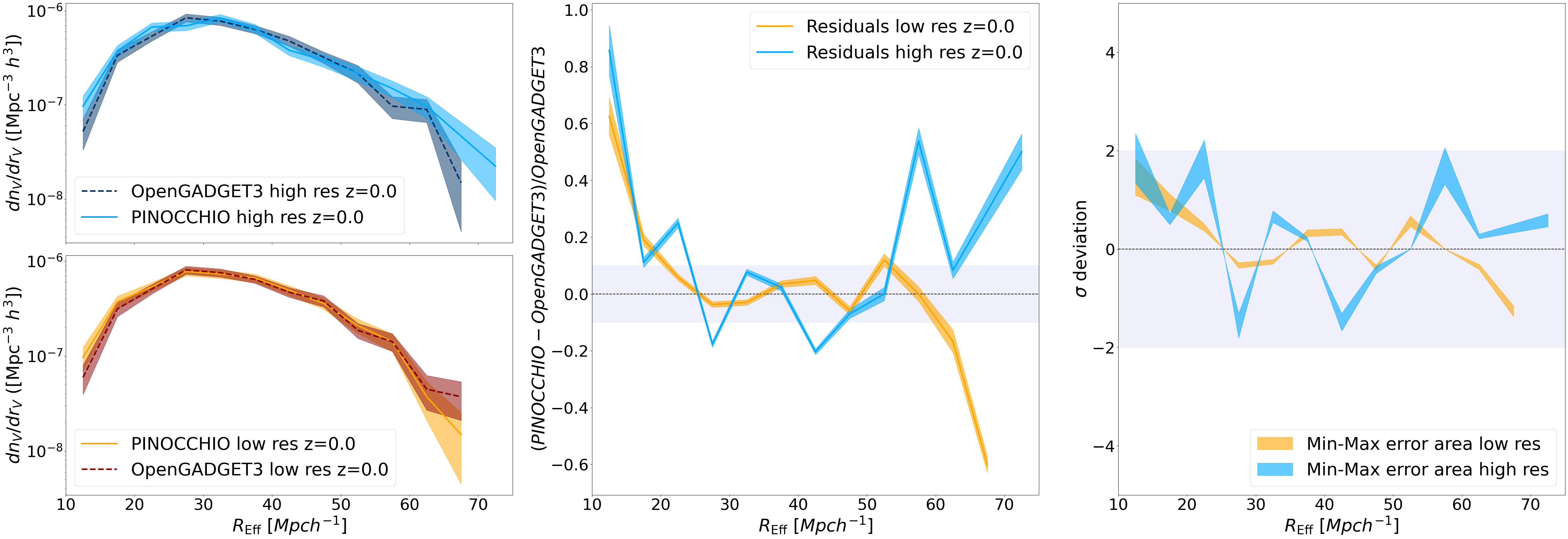}
    \caption{Same as Figure \ref{HMF_comparison}, but for the VSFs.}
    \label{VSF_comparison}
\end{figure*}

Figure \ref{VSF_comparison} presents the same set of measurements as in Figure~\ref{HMF_comparison}, but for the VSFs derived from \og\ and \PINOCCHIO. The left panels show that the two codes agree well for both high and low resolution at z = 0. Both \PINOCCHIO\ and \og\ yield VSFs that have qualitatively the same shape. The two codes agree well around the peak of the distributions, with larger discrepancies at the low- and high-size tails. This is verified by the middle panels, which demonstrate that \PINOCCHIO\ remains within the $\pm 10\%$ range of \og\ for $25 {\rm Mpc}/h < R_{\mathrm{eff}} < 55 {\rm Mpc}/h$ (corresponding to $\sim 73 \%$ of the total number of voids at both resolutions), consistent with the HMFs comparisons discussed in Section \ref{HMF_comparison_section}. As in the HMF comparison, the right column shows that most values fall within $\pm2\sigma$, suggesting that the observed deviations are consistent with statistical fluctuations (even though 20–80\% discrepancies may appear large), with no evidence of a systematic bias across the $R_{\mathrm{eff}}$ range.

Even though the halo catalogs are matched in number density, as described in Section ~\ref{HMF_comparison_section}, the clustering of halos in \PINOCCHIO\ has less power at small scales compared to \og. Since \vide\, estimates the density field using a Voronoi tessellation, differences in tracer clustering can influence void identification and subsequent void properties. In particular, the weaker clustering of \PINOCCHIO\ halos leads to a coarser density field, which alters how voids are detected and classified. 

\begin{figure}[t]
    \centering
    \includegraphics[width=0.95\columnwidth]
    {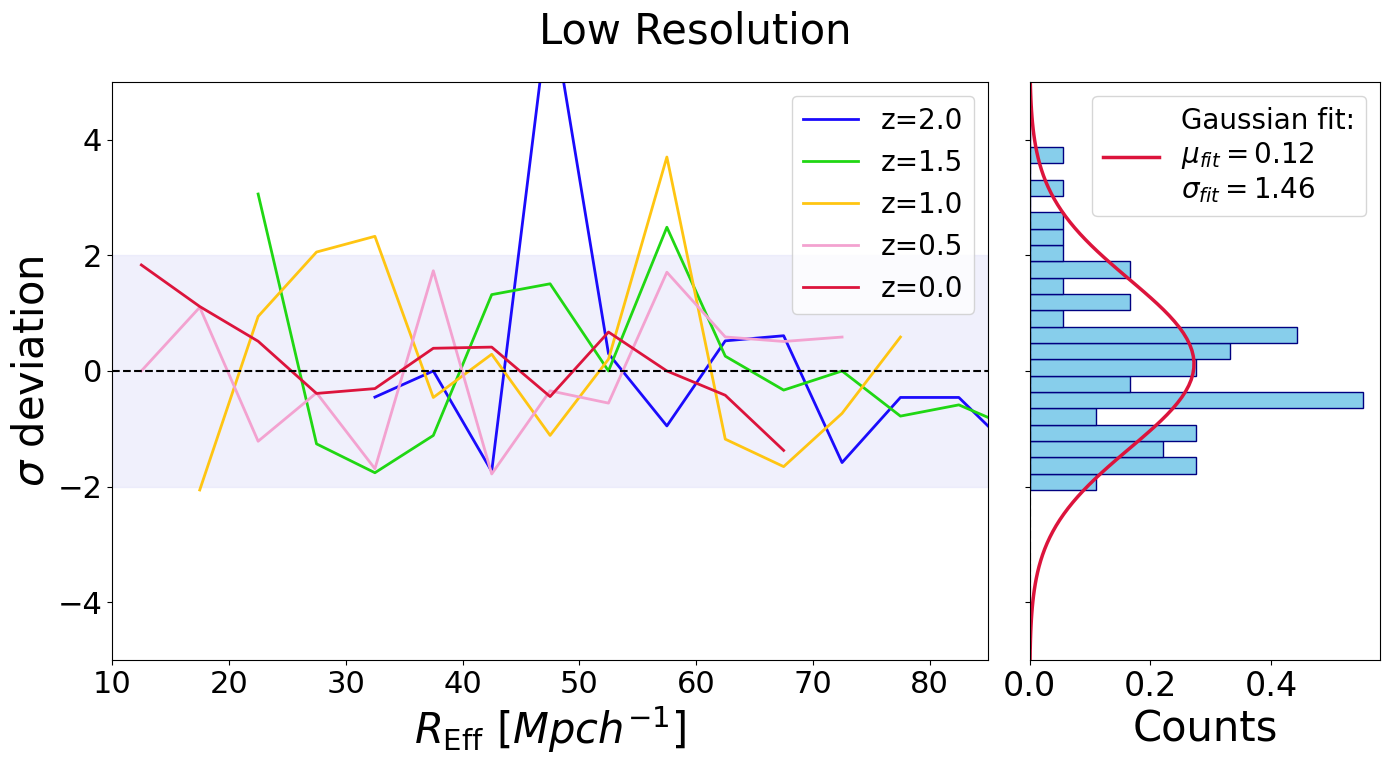}

    \vspace{0.1em}
    
    \includegraphics[width=0.95\columnwidth]{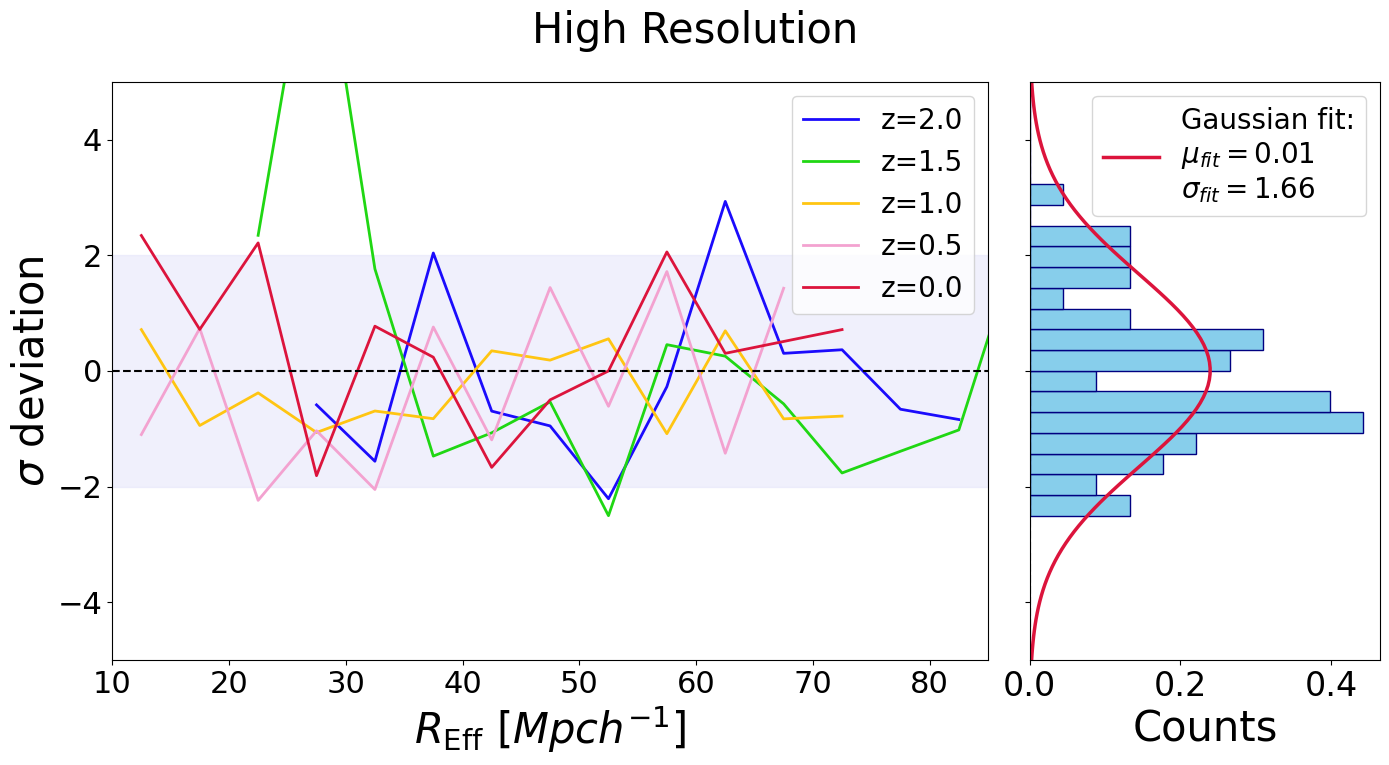}
    \caption{Same as Figure \ref{HMF_evolution}, but for the VSFs as a function of $R_{\mathrm{Eff}}$. }
    \label{VSF_evolution}
\end{figure} 
This reduced small-scale power directly affects the VSF. In \PINOCCHIO, weaker nonlinear clustering suppresses halo collapse and merging relative to \og, allowing more small voids in the void-in-cloud regime to survive rather than being erased by environmental collapse. For larger voids, the lack of small scale overdensities in \PINOCCHIO\ can reduce the fragmentation of large underdense regions, potentially leading to an an excess of large voids relative to \og. At lower resolution, however, the under-resolved density field can lead to spurious overdense bridges that artificially split or isolate underdense regions, suppressing the formation of large voids through void-in-void merging. The improved agreement between the codes for the largest voids in low-resolution simulations likely reflects a regime where both are similarly limited in resolving small-scale structure. 
\begin{figure*}[t]
    \centering
    \includegraphics[width=0.97\textwidth]{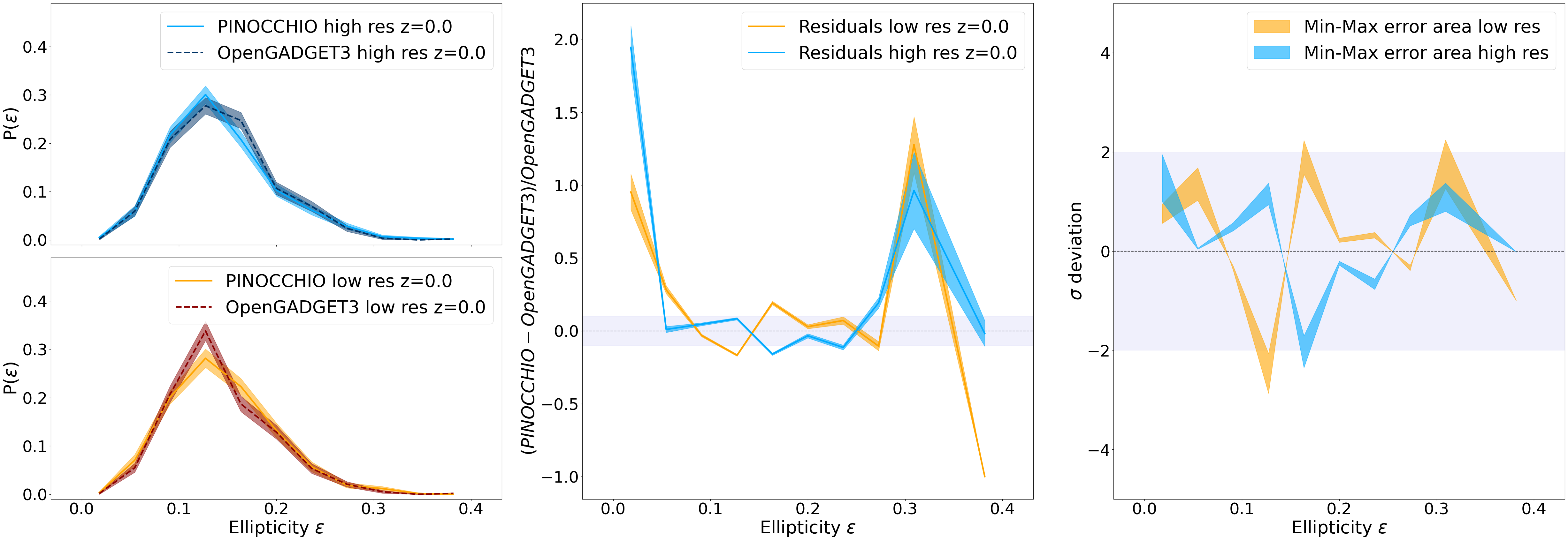}
    \caption{Same as Figure \ref{HMF_comparison}, but for the VEFs.}
    \label{Ellipticity_comparison}
\end{figure*}
Additionally, differences in resolution and tracer populations further contribute to the observed discrepancies. At high resolution, \PINOCCHIO\ shows an excess of large voids relative to \og, possibly due to its improved ability to resolve intermediate-mass halos and thereby alter the reconstructed density field. As shown in Appendix~\ref{Resolution_comparison}, the VSFs in \PINOCCHIO\ vary more strongly with resolution than in \og, which remains largely stable. This suggests a stronger resolution dependence in \PINOCCHIO, although the statistical significance of this trend remains limited.

Figure \ref{VSF_evolution} repeats the measurements from Figure \ref{HMF_evolution} but for the VSFs at multiple resdshifts. These results confirm the robustness of the \PINOCCHIO\ method compared to \og\ across cosmic time. For both resolutions, the agreement remains well within the $\pm 2\sigma$ shaded region for most void size scales, demonstrating reliable performance across a wide range of epochs. As for the HMF, the right panels show that in both resolution cases, the fitted distributions show $|\mu_{fit}| - \sigma_{fit} < 0$, implying that residuals are symmetrically distributed around zero and no significant bias is present, when marginalizing over redshift. This indicates that the observed deviations are primarily statistical in nature and could potentially be reduced with improved measurements, such as larger sample size by increasing the resolution and/or box size. 

\subsection{Void ellipticity function}
Figure \ref{Ellipticity_comparison} presents the same set of measurements as in Figure \ref{HMF_comparison}, but for the VEFs. 
The left panels highlight a reasonable agreement between the two methods. For both low- and high-resolution simulations, the four curves, along with their respective error bars, exhibit substantial overlap, with a stronger peak in the ellipticity distribution given by \PINOCCHIO\ in the high resolution case. 
The middle panels indicate that \PINOCCHIO\ generally follows the \og\ ellipticity distribution within the $\pm 10\%$ deviation range over the $0.05 < \epsilon < 0.25$, though residuals show noticeable noise, preventing a precise one-to-one correspondence across all bins. The deviations become larger at the extremes of the ellipticity range, which can be attributed to the small number of voids with such shapes and the resulting shot noise. Values of ($\epsilon>0.4$) are excluded, as void counts in these bins are insufficient to draw statistically meaningful conclusions. As in the VSFs case, the right panels show that most differences fall within $\pm 2\sigma$, consistent with statistical fluctuations and show no clear sign of systematic bias across the $\epsilon$ range. Resolution effects appear modest, with broadly similar distributions across the two resolutions, although the higher-resolution case shows slightly better agreement within the $\pm 2\sigma$ range, likely due to the higher resolution tracer sample giving a more robust sampling of the of the underlying density field. 
\begin{figure}[t]
    \centering
    \includegraphics[width=0.95\columnwidth]
    {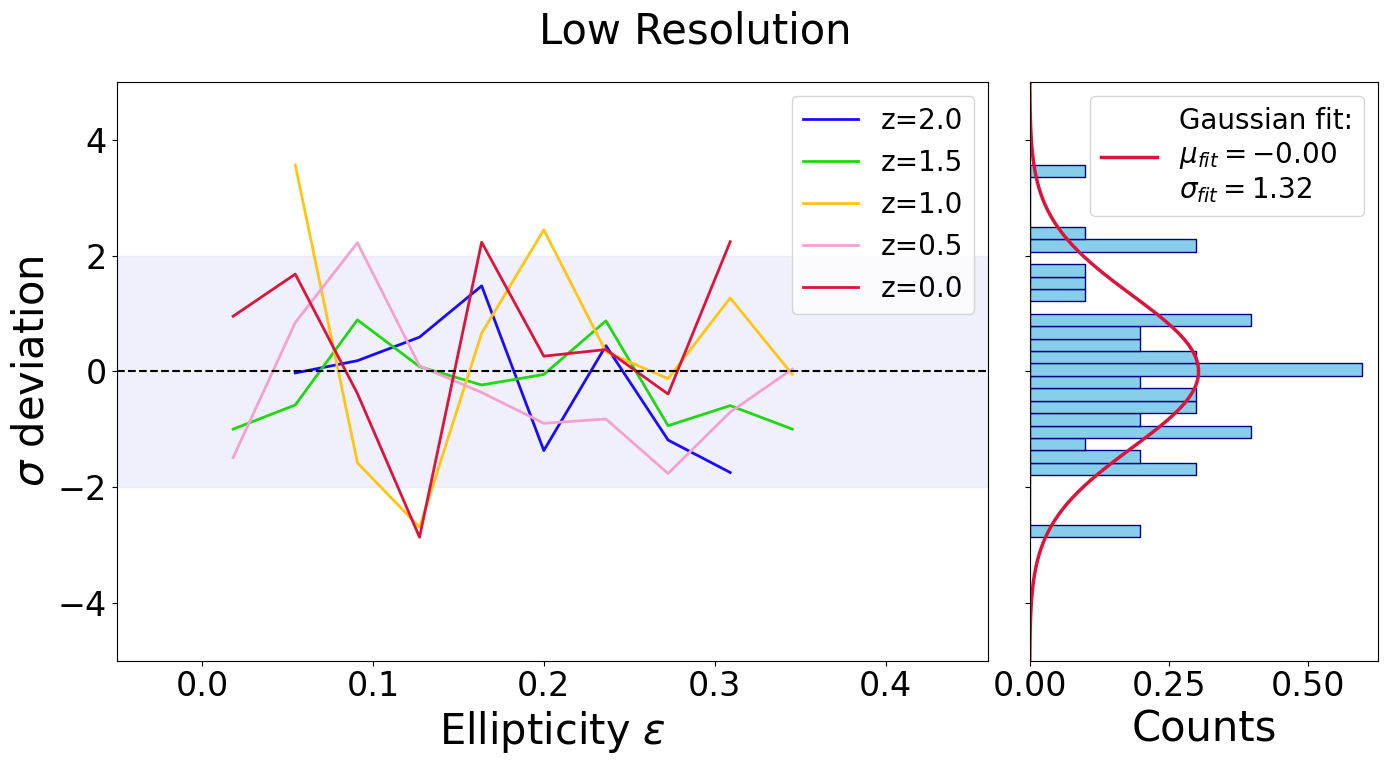}

    \vspace{0.1em}
    
    \includegraphics[width=0.95\columnwidth]{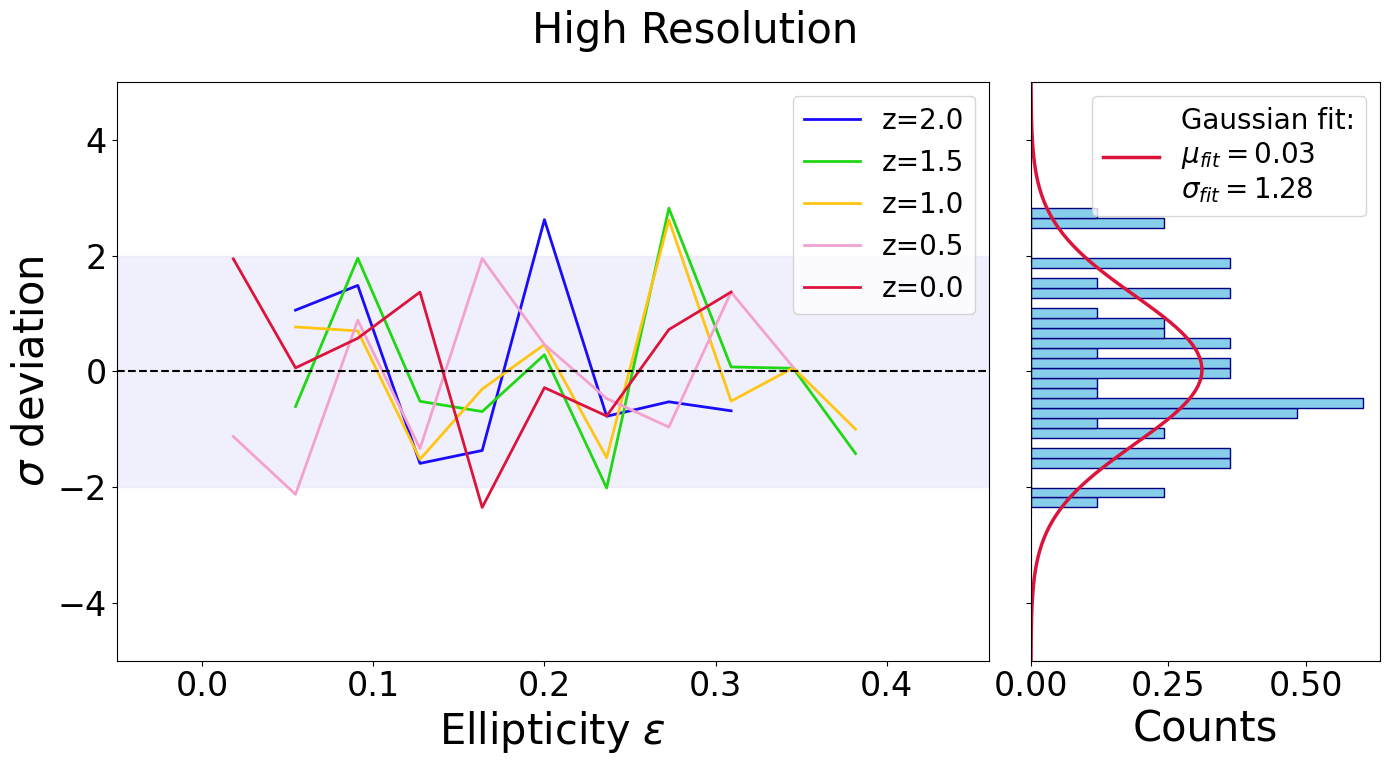}
    \caption{Same as Figure \ref{HMF_evolution}, but for the VEFs as a function of $\epsilon$.}
    \label{Ellipticity_evolution}
\end{figure}

Figure \ref{Ellipticity_evolution} extends the comparison of VEFs to multiple redshifts, following the same methodology applied in Figure \ref{HMF_evolution}. The results confirm that the ellipticity distributions from \PINOCCHIO\ and \og\ remain statistically consistent as a function of redshift. As observed for the VSFs, the differences between the two methods lie within the $\pm 2 \sigma$ range for most ellipticity values at all redshifts, suggesting no significant systematics or biases introduced by the \PINOCCHIO\ method. The Gaussian fits in the right panels further support this, showing a slightly better agreement relative to the VSF case, with $\mu_{\mathrm{fit}} \approx 0$ and $\sigma_{\mathrm{fit}} \approx 1.3$ for both resolutions.
\begin{figure*}[t]
    \centering
    \includegraphics[width=0.97\textwidth]{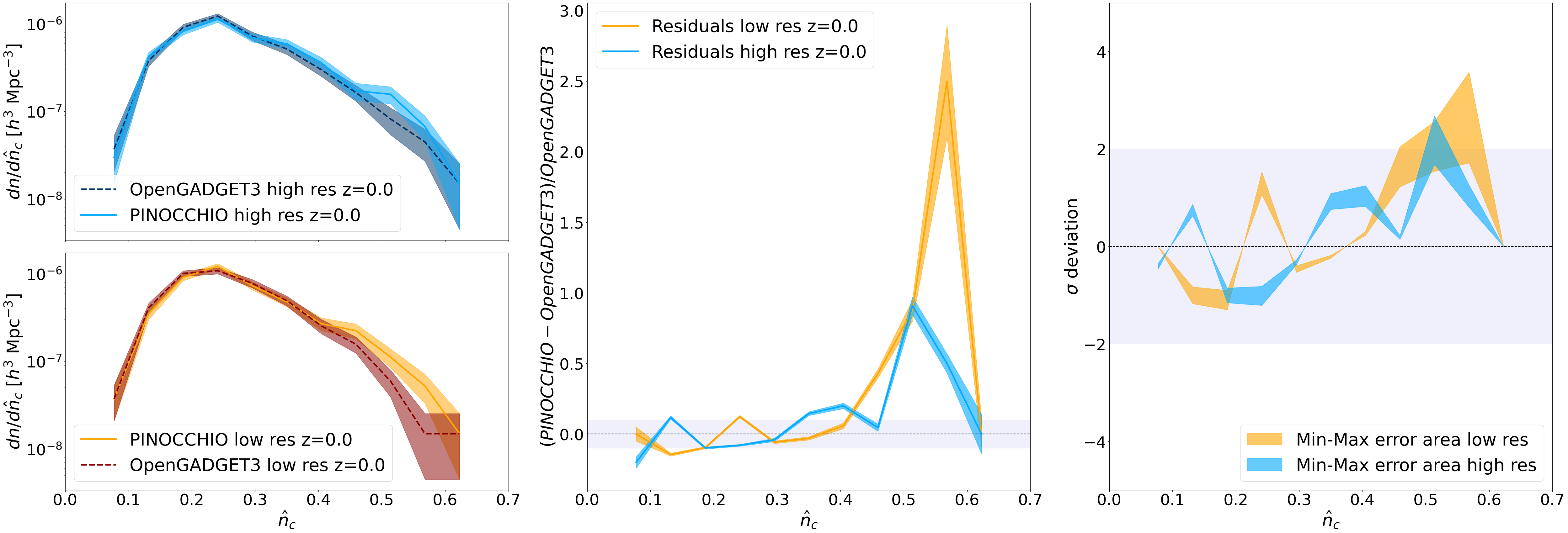}
    \caption{Same as Figure \ref{HMF_comparison}, but for the CDFs.}
    \label{CoreDens_comparison}
\end{figure*}
Interestingly, the evolution of the  VEFs $\sigma$ deviations with redshift reveals that, even as voids tend to become slightly more elliptical at earlier epochs due to stronger tidal interactions, \PINOCCHIO\ effectively captures this trend in alignment with \og. This consistency demonstrates the reliability of \PINOCCHIO\ in reproducing the structural and dynamical properties of voids across cosmic time, despite the differences in the underlying halo distributions.

\subsection{Core density function}
Figure \ref{CoreDens_comparison} presents the same set of measurements as in Figure \ref{HMF_comparison}, but for the CDFs derived from \og\ and \PINOCCHIO. The comparison in the left panels reveals a generally good agreement at low core densities $\hat{n}_{C}$, where both simulations exhibit similar void properties. Since the core densities $\hat{n}_{C}$, as defined in Eq. \eqref{core_density}, corresponds to the region of lowest density within a void, the observed behavior suggests that \PINOCCHIO\ performs more accurately in underdense regions compared to overdense regions. This trend is confirmed in the middle panels: \PINOCCHIO\ generally follows the \og\ core density distribution within the $\pm 10\%$ deviation range over the interval $\sim 0.05 < \hat{n}_{C} < 0.35-0.4$. The distribution in the right column reinforces the trend seen in previous statistics: the majority of values remain within $\pm2\sigma$, indicating that differences between \PINOCCHIO\ and \og\ are largely driven by statistical fluctuations. Notably, across the full $\hat{n}_{C}$ range, there is no clear evidence of a systematic bias, further validating the consistency of \PINOCCHIO\ in capturing the low-density interiors of voids.
\begin{figure}[t]
    \centering
    \includegraphics[width=0.95\columnwidth]{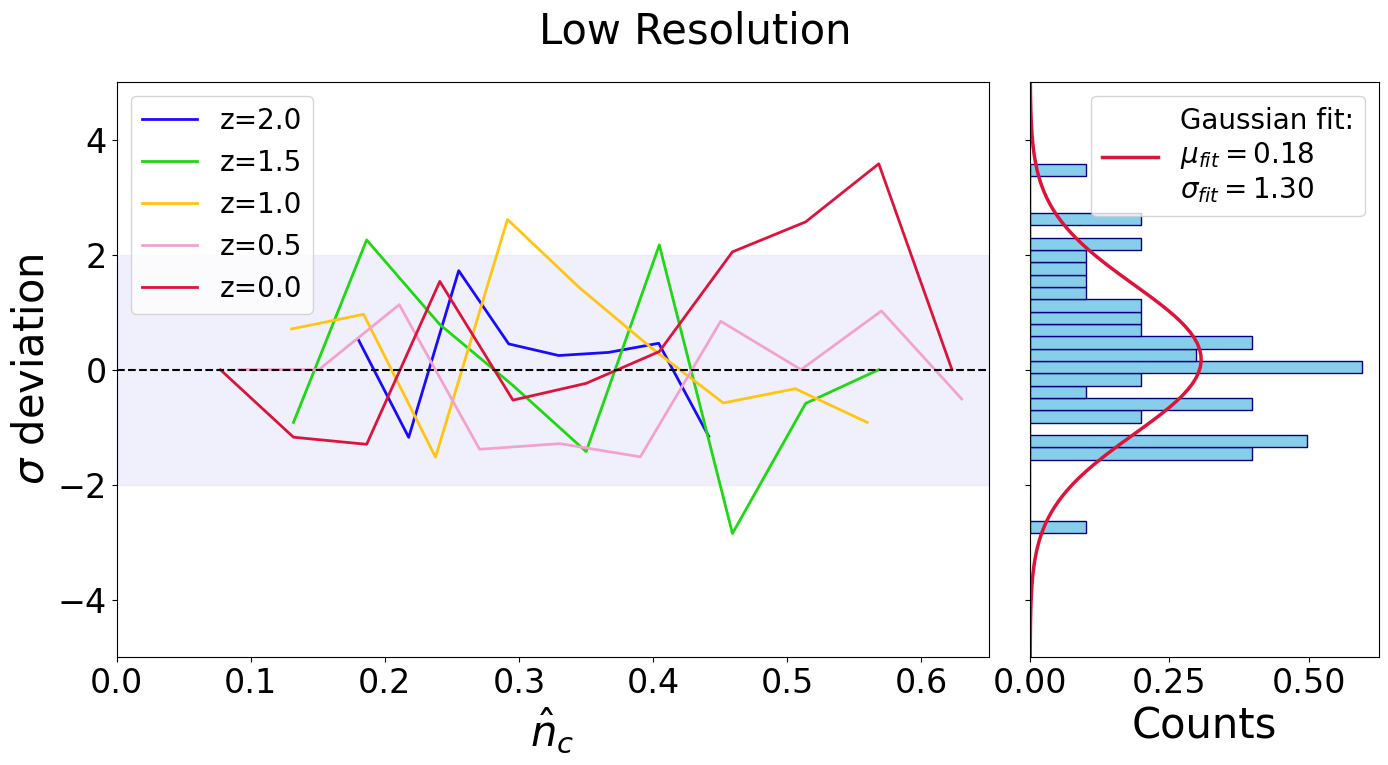}

     \vspace{0.1em}
     
    \includegraphics[width=0.95\columnwidth]{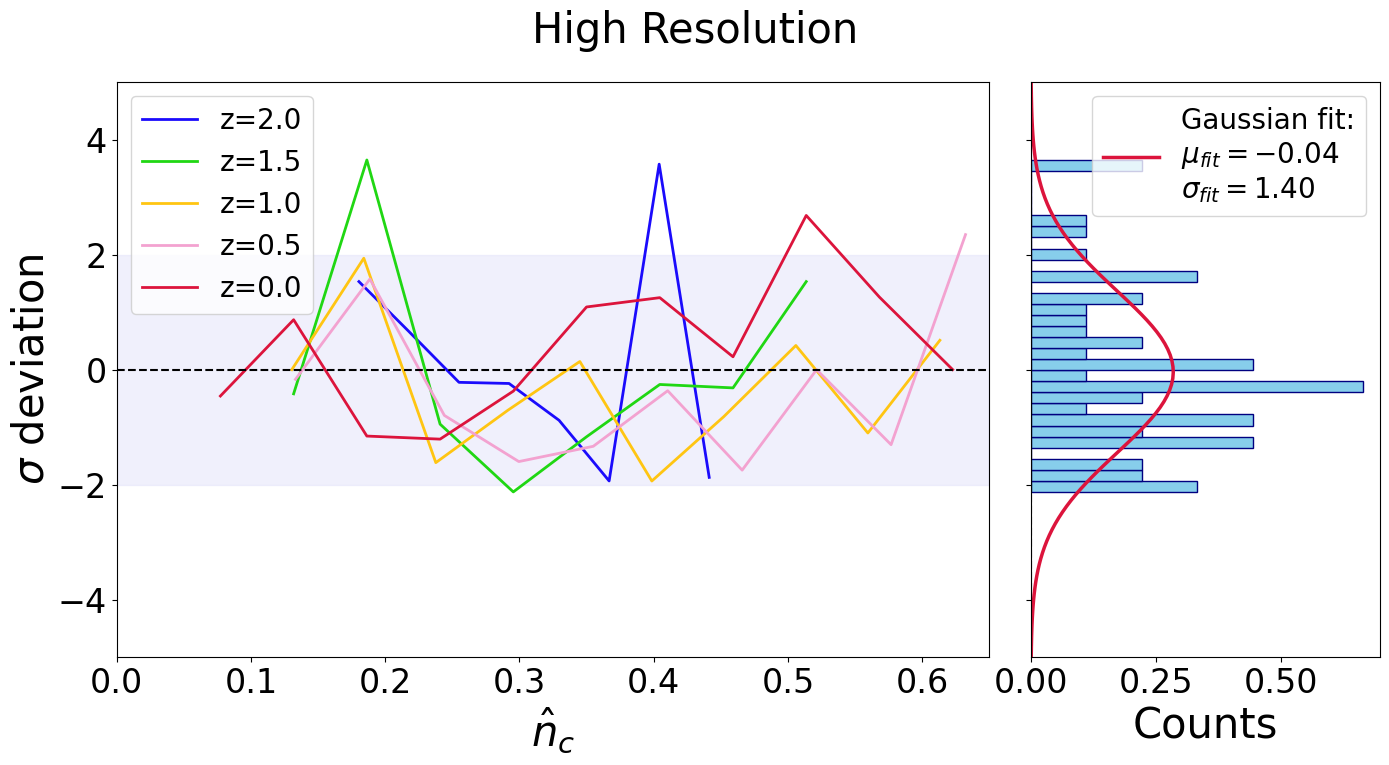}
    \caption{Same as Figure \ref{HMF_evolution}, but for the CDFs as a function of $\hat{n}_{C}$.}
    \label{CoreDens_evolution}
\end{figure}

At higher core densities, a systematic overestimation of void numbers is observed in \PINOCCHIO. Unlike the other summary statistics, the CDF appears more challenging for \PINOCCHIO\ to reproduce accurately, though the overall agreement remains within $2\sigma$. We argue that the \frag\ algorithm in \PINOCCHIO\ (Section \ref{PINOCCHIO}), may boost the relative number of low-mass halos, as can also be seen from the HMF comparison in Figure \ref{HMF_comparison}, which in turn enhances the abundance of voids with higher core densities. This interpretation is further supported by the resolution dependence of the discrepancy: better agreement is seen at higher resolution, where \frag\ is expected to operate more effectively. These trends are illustrated in Appendix ~\ref{Resolution_comparison}, where the resolution dependence of the CDFs is explored in more detail.

Figure~\ref{CoreDens_evolution} shows the redshift evolution of the CDF comparison, applying the same approach as in Figure~\ref{HMF_evolution}. Across all epochs, the agreement between \PINOCCHIO\ and \og\ remains stable, with most deviations falling within the $\pm 2\sigma$ range. The fitted Gaussian distributions in the right panels further suggest that these residuals are statistically distributed around zero, showing no evidence of redshift-dependent biases. This consistency supports the robustness of \PINOCCHIO\ in tracing the evolution of the most underdense regions of the cosmic web. Since such regions are expected to feel the effects of accelerated cosmic expansion earlier in time, the agreement with \og\ suggests that \PINOCCHIO\ effectively captures this behavior in line with full $N$-body dynamics.
                                            
\subsection{Radial density profile}
\begin{figure*}[t]
    \centering
    \includegraphics[width=0.97\textwidth]{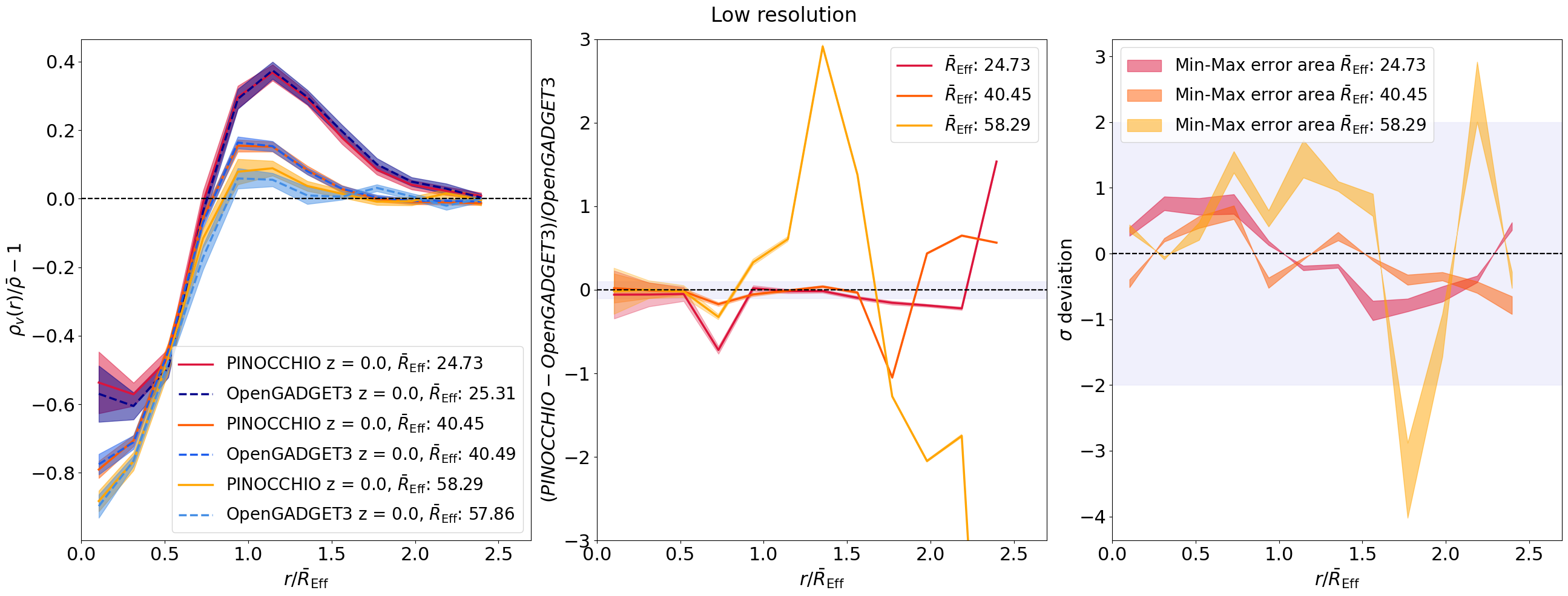}

    \vspace{0.1em} 

    \includegraphics[width=0.97\textwidth]{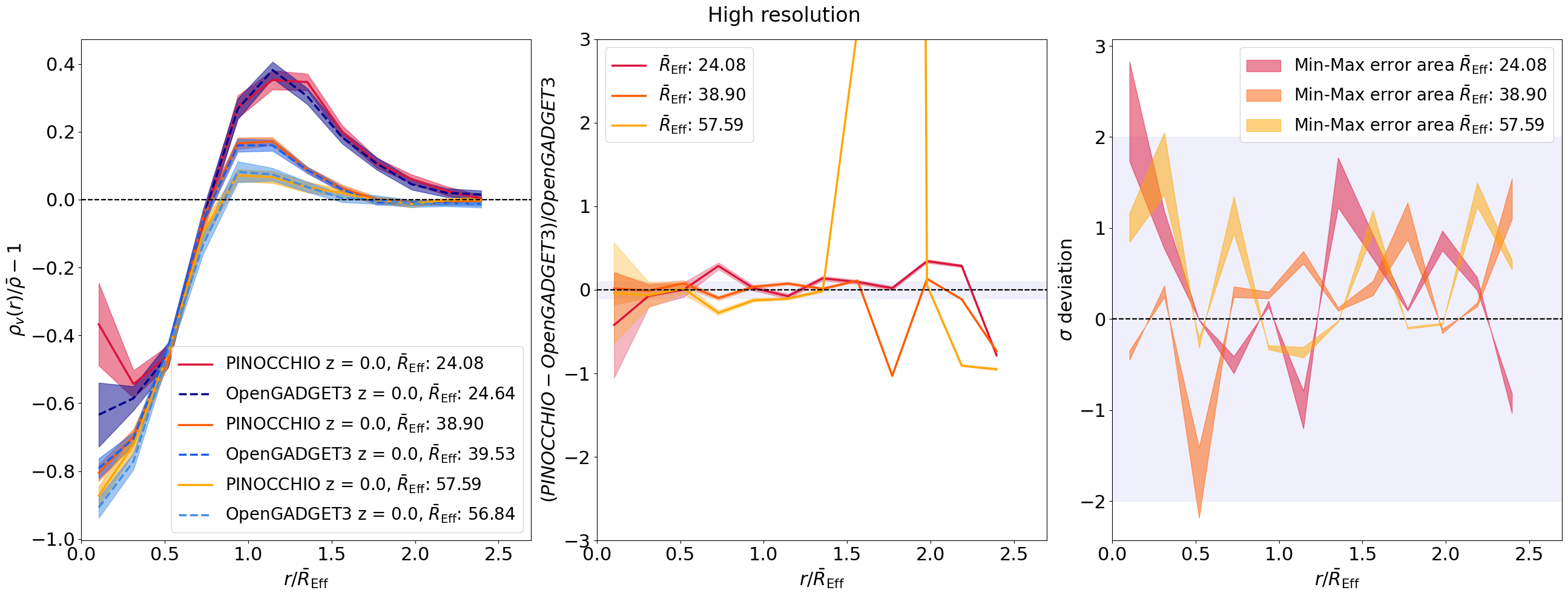}
    
    \caption{Same as Figure \ref{HMF_comparison}, but for the RDPs. The left panels show the stacked radial density profiles, with the legend indicating the mean  $R_{\mathrm{eff}}$ in each of the three linear bins used for stacking. For clarity, only the mean $\bar{R}_{\mathrm{eff}}$ from the \PINOCCHIO\ are reported in the middle and right panels. }
    \label{Profile_comparison}
\end{figure*}
Figure \ref{Profile_comparison} presents the same type of measurement depicted in Figure \ref{HMF_comparison}, but for the RDPs derived from \og\ and \PINOCCHIO\ after applying the stacking procedure described in Section~\ref{VOID_statistics}. The stacking bins are defined using the range of effective void radii ${R}_{\mathrm{eff}}$ from the \PINOCCHIO\ simulation. This range is then applied consistently to both simulations, ensuring that the void stacking and jackknife resampling are performed over a common reference scale. The average void radii for each bin are indicated in the legend of Figure~\ref{Profile_comparison}, and the bin edges along with the number of voids per simulation are summarized in Table~\ref{tab:void_bins}. 

The left panels show that the RDPs, along with their respective error bars, exhibit good overlap across all three linear bins. Although the bin edges are defined based on the \PINOCCHIO\ void size range, the average $R_{\mathrm{eff}}$ values, indicated in the legend, are closely matched, suggesting that the void populations in \og\ and \PINOCCHIO\ are very similar. As expected, the stacked void profiles are deeply underdense at their centers, with the central density generally decreasing as the void size increases. The rise in central density for the smallest voids is due to the limited number of tracers at scales below the mean separation, which can bias the estimate from Eq.~\eqref{density_in_shell}, particularly when a tracer falls within a small central shell~\citep{Schuster:2022ogh}. 
\begin{table}[t]
\centering
\caption{Number of voids per radial bin.}
\label{tab:void_bins}
\resizebox{0.98\columnwidth}{!}{%
\begin{tabular}{cc|cc|cc}
\toprule
\multicolumn{2}{c|}{$R_{\mathrm{eff}}$ range [Mpc/$h$]} & \multicolumn{2}{c|}{\PINOCCHIO} & \multicolumn{2}{c}{\og} \\
Low res & High res & Low res & High res & Low res & High res \\
\midrule
$[12.08, 32.22]$ & $[11.82, 31.47]$ & 285 & 277 & 271 & 265 \\
$[32.22, 52.36]$ & $[31.47, 51.13]$ & 270 & 267 & 267 & 277 \\
$[52.36, 72.50]$ & $[51.13, 70.78]$ &  39 &  55 &  41 &  47 \\
\bottomrule
\end{tabular}%
}
\tablefoot{The first column lists the bin edges for both low- and high-resolution runs, defined from the \PINOCCHIO\ ${R}_{\mathrm{eff}}$ distribution. The second and third columns report the number of voids from the \PINOCCHIO\ and \og\ simulations, respectively, at redshift z = 0.}
\end{table}

In general, all profiles show clearly defined compensation walls around $r = R_{\mathrm{eff}}$, a typical feature of voids identified via the watershed method used in \vide. As expected, the profiles converge toward the mean background density at sufficiently large distances from the void center. The middle panel demonstrates that \PINOCCHIO\ generally tracks the \og\ profiles within the $\pm 10\%$ deviation range in the inner void regions, up to $r/R_{\mathrm{eff}} \sim 0.5$. The agreement is better for smaller voids, while the larger bins show increased noise due to lower void numbers as noted also for the VSF in Figure~\ref{VSF_comparison} . Additionally, the overall consistency between the two codes appears to improve with higher resolution, indicating that increased tracer density enhances the robustness of the RDP measurements.
The results in the right column remain mostly within the $\pm2\sigma$ interval, reinforcing the interpretation that the differences are dominated by statistical fluctuations rather than persistent biases across the $r/\bar{R}_{\mathrm{eff}}$ range.
\begin{figure*}[t]
    \centering
    \includegraphics[width=0.95\textwidth]{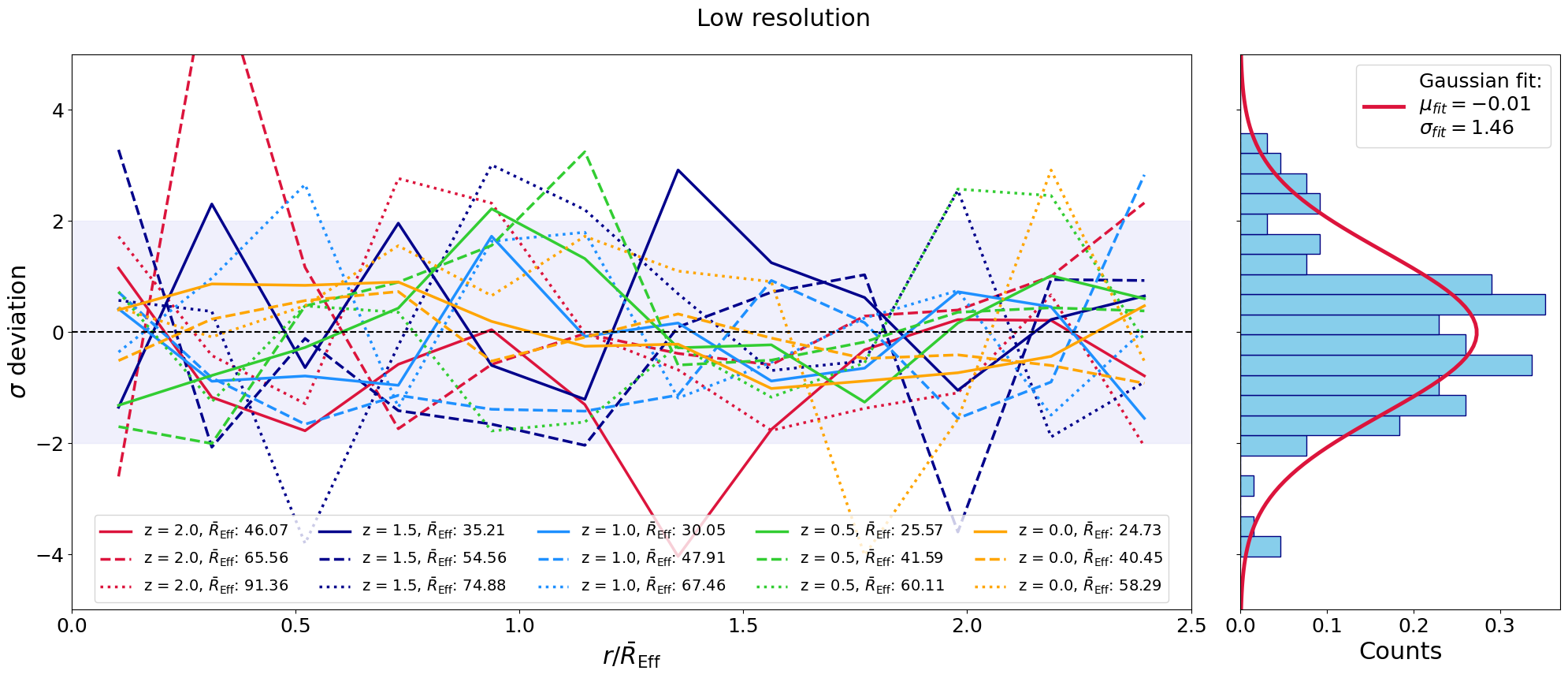}

    \vspace{0.1em}
    
    \includegraphics[width=0.95\textwidth]{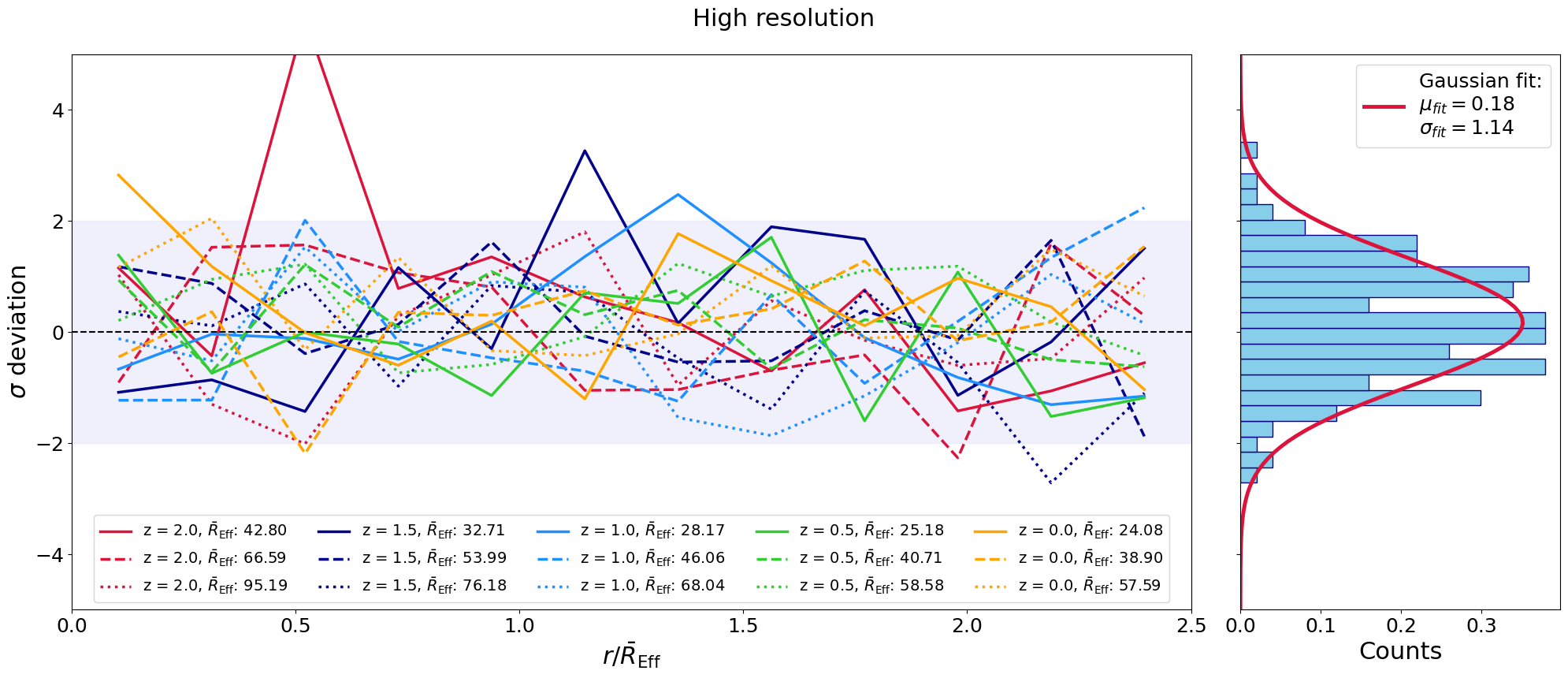}

    \caption{Left panels: same as Figure \ref{HMF_evolution}, but for the RDPs as a function of $r/\bar{R}_{\mathrm{eff}}$. The different stacking bins and redshifts are color-coded. Right panels: overall distribution of the measurements in the left panels, aggregated over all redshifts and size bins, with overlaid Gaussian fits.}
    \label{Profile_evolution}
\end{figure*}

Figure~\ref{Profile_evolution} extends the comparison of RDPs to multiple redshifts. It is organized in two rows and two columns: the top panels correspond to the low-resolution simulations ($N_p = 512^3$), while the bottom panels present results from the high-resolution case ($N_p = 1024^3$). The left panels replicate the same measurements of Figure~\ref{HMF_evolution}, for the stacked RDPs for each of the three linear bins across all redshifts. The right panels display the overall distribution of these deviations, aggregated over all redshifts and size bins, with overlaid Gaussian fits. The results confirm that the RDPs from \PINOCCHIO\ and \og\ remain statistically consistent over cosmic time. Similar to the VSFs, VEFs, and CDFs, the deviations in the left panels fall within the $\pm 2 \sigma$ range with no redshift trend. The Gaussian fit in the right panels provide a complementary statistical check: the relative difference normalized by the \og\ errors (Eq.~\eqref{eq:errors_max}) are centered near zero and exhibit a spread comparable to, or slightly above, the expected statistical noise. This behavior reinforces the view that differences between the two methods arise from random variation rather than systematic bias. 

The agreement improves with resolution, reflecting better convergence in the tracer distribution. These findings reinforce the reliability of \PINOCCHIO\ in capturing the large-scale structure around voids. This is especially relevant for studies aiming to extract cosmological information from void environments, including potential constraints on modified gravity or dark energy models through the shape and depth of the radial density profiles.

\section{Conclusions}
\label{Conclusions}
In this work, we have assessed the accuracy of the semi-analytic code \PINOCCHIO\ in reproducing void summary statistics by comparing it with the full $N$-body code \og. To ensure a fair comparison of the void properties, we used identical ICs and constructed halo catalogs matched in number density, thereby ensuring statistically comparable tracer distributions.

By applying the watershed void finder \vide\ to the halo catalogs, we investigated four key void summary statistics: the void size function (VSF), void ellipticity function (VEF), core density function (CDF), and radial density profiles (RDP). Across all metrics, \PINOCCHIO\ demonstrated excellent agreement with \og\ within statistical uncertainties, with no evidence of systematic biases. In particular:
\begin{itemize}
    \item VSF: \PINOCCHIO\ reproduces the void abundance as a function of size across redshift and resolution (Figures~\ref{VSF_comparison} and~\ref{VSF_evolution}), with a level of agreement comparable to that achieved for HMF in overdense regions (Figures~\ref{HMF_comparison} and~\ref{HMF_evolution}). Differences at large void sizes are attributed to resolution-dependent halo clustering effects.
    \item VEF: the ellipticity distributions show excellent consistency across multiple redshifts, suggesting that \PINOCCHIO\ captures the large-scale tidal influence on void shapes (Figures~\ref{Ellipticity_comparison} and~\ref{Ellipticity_evolution}).
    \item CDF: \PINOCCHIO\ performs particularly well at low core densities, where its LPT-based approach is expected to be most accurate (Figures~\ref{CoreDens_comparison} and~\ref{CoreDens_evolution}). Deviations at higher core densities are mild and resolution-dependent, likely related to \frag\ and small-halo statistics.
    \item RDP: the stacked radial density profiles agree well across all redshifts, reinforcing \PINOCCHIO’s reliability in reproducing the structure, environment, and evolution of voids (Figures~\ref{Profile_comparison} and~\ref{Profile_evolution}).
\end{itemize}

Across all these statistics, the differences between \PINOCCHIO\ and \og\ results remain mostly within $\pm10\%$ variation range and below the $2\sigma$ deviation level. Because the two simulations share the same ICs, their statistical uncertainties are correlated; although a full covariance estimate is beyond scope, we bracketed the significance with conservative estimators that provide upper and lower limits on the true level of statistical disagreement. Additional tests on a larger box, with an order-of-magnitude more voids, confirmed consistent results (Appendix~\ref{Large_Box}), showing that our conclusions are robust against sample variance and finite-box effects.

Overall these results demonstrate that \PINOCCHIO\ reproduces void summary statistics with good accuracy, particularly in the underdense regime where linear and quasi-linear dynamics dominate. This makes it a promising tool for cosmological applications involving voids, especially in the context of upcoming large-volume surveys such as Euclid and LSST, where computational efficiency and statistical robustness are crucial. A natural valuable next step would be to test the performance of \PINOCCHIO\ in a lightcone geometry, incorporating observational systematics, to assess its direct applicability to survey data. 

These results can also be interpreted as quantifying the bias that \PINOCCHIO\ introduces in void statistics. Once characterized, such a bias can be incorporated into data analyses to recover unbiased constraints. Furthermore, a $\sim$10\% uncertainty in the statistic is not necessarily limiting for parameter inference, since the steepness of both the HMF and VSF often makes such effects subdominant. The impact may also depend on tracer density, with small voids in dense surveys (e.g. DESI) being harder to model, while larger voids should remain robustly captured. Moreover such biases are not expected to vary significantly with cosmology within $\Lambda$CDM, and previous works~\citep{Munari:2016aut} have shown that the calibration of \PINOCCHIO\ is cosmology-independent. Applications to modified gravity would require re-calibration and are left for future work.

Since \PINOCCHIO\ is already widely used for generating large ensembles of simulations to estimate and validate covariance matrices, a natural extension of this work would be to investigate whether \PINOCCHIO\ can reliably reproduce the covariances of void statistics. The good agreement found here suggests this may be the case, as already demonstrated for halo statistics. Validating the precision of void covariance estimation would be an important step toward their use in high-precision cosmological inference from observational data, while taking full advantage of \PINOCCHIO’s computational efficiency.

More broadly, the accuracy of \PINOCCHIO\ in reproducing void statistics, particularly in underdense regimes, makes it well suited for survey forecasts and theoretical modeling, including tests of modified gravity and dark energy. Future work will explore its performance in non-standard cosmologies and its direct application to cosmological inference from void statistics.

\begin{acknowledgements}
We would like to thank Alessandra Fumagalli and Sofia Contarini for valuable discussions, and Daniele Tavagnacco for technical support with the void finder on a dedicated machine at the OATs HPC system. This paper is supported by the Fondazione ICSC, Spoke 3 Astrophysics and Cosmos Observations. National Recovery and Resilience Plan (Piano Nazionale di Ripresa e Resilienza, PNRR) Project ID CN\_00000013 `Italian Research Center on High-Performance Computing, Big Data and Quantum Computing' funded by MUR Missione 4 Componente 2 Investimento 1.4: Potenziamento strutture di ricerca e creazione di `campioni nazionali di R\&S (M4C2-19 )' - Next Generation EU (NGEU). PM acknowledges  support by the PRIN 2022 PNRR project (code no. P202259YAF) funded
by “European Union – Next Generation EU”, Mission 4, Component 1, CUP
J53D23019100001. We acknowledge support from the LMU Faculty of Physics in Munich. NS acknowledges support from the french government under the France 2030 investment plan, as part of the Initiative d’Excellence d’Aix-Marseille Universit\'e - A*MIDEX AMX-22-CEI-03. 
\end{acknowledgements}

\bibliographystyle{aa}
\bibliography{biblio}

\begin{thebibliography}{55}
\expandafter\ifx\csname natexlab\endcsname\relax\def\natexlab#1{#1}\fi

\bibitem[{Ade {et~al.}(2016)}]{Planck:2015fie}
Ade, P. A.~R. {et~al.} 2016, A\&A, 594, A13

\bibitem[{Alcock \& Paczynski(1979)}]{Alcock:1979mp}
Alcock, C. \& Paczynski, B. 1979, Nature, 281, 358

\bibitem[{Barnes \& Hut(1986)}]{Barnes:1986nb}
Barnes, J. \& Hut, P. 1986, Nature, 324, 446

\bibitem[{Bond {et~al.}(1991)Bond, Cole, Efstathiou, \& Kaiser}]{Bond:1990iw}
Bond, J.~R., Cole, S., Efstathiou, G., \& Kaiser, N. 1991, ApJ, 379, 440

\bibitem[{Bos {et~al.}(2012)Bos, van~de Weygaert, Dolag, \& Pettorino}]{Bos:2012wq}
Bos, E. G.~P., van~de Weygaert, R., Dolag, K., \& Pettorino, V. 2012, MNRAS, 426, 440

\bibitem[{Castro {et~al.}(2023)}]{Euclid:2022dbc}
Castro, T. {et~al.} 2023, A\&A, 671, A100

\bibitem[{Contarini {et~al.}(2021)Contarini, Marulli, Moscardini, Veropalumbo, Giocoli, \& Baldi}]{Contarini:2020fdu}
Contarini, S., Marulli, F., Moscardini, L., {et~al.} 2021, MNRAS, 504, 5021

\bibitem[{Contarini {et~al.}(2023)Contarini, Pisani, Hamaus, Marulli, Moscardini, \& Baldi}]{Contarini:2022mtu}
Contarini, S., Pisani, A., Hamaus, N., {et~al.} 2023, ApJ, 953, 46

\bibitem[{Contarini {et~al.}(2019)Contarini, Ronconi, Marulli, Moscardini, Veropalumbo, \& Baldi}]{Contarini:2019qwf}
Contarini, S., Ronconi, T., Marulli, F., {et~al.} 2019, MNRAS, 488, 3526

\bibitem[{{Davies} {et~al.}(2021){Davies}, {Cautun}, {Giblin}, {Li}, {Harnois-D{\'e}raps}, \& {Cai}}]{Davies2021}
{Davies}, C.~T., {Cautun}, M., {Giblin}, B., {et~al.} 2021, \mnras, 507, 2267

\bibitem[{{Davies} {et~al.}(2019){Davies}, {Cautun}, \& {Li}}]{Davies2019}
{Davies}, C.~T., {Cautun}, M., \& {Li}, B. 2019, \mnras, 490, 4907

\bibitem[{Dolag {et~al.}(2009)Dolag, Borgani, Murante, \& Springel}]{Dolag:2008ar}
Dolag, K., Borgani, S., Murante, G., \& Springel, V. 2009, MNRAS, 399, 497

\bibitem[{Duncan {et~al.}(1998)Duncan, Levison, \& Lee}]{Duncan_1998}
Duncan, M.~J., Levison, H.~F., \& Lee, M.~H. 1998, AJ, 116, 2067

\bibitem[{Efstathiou {et~al.}(1985)Efstathiou, Davis, Frenk, \& White}]{Efstathiou:1985re}
Efstathiou, G., Davis, M., Frenk, C.~S., \& White, S. D.~M. 1985, ApJS, 57, 241

\bibitem[{Fumagalli {et~al.}(2025)Fumagalli, Castro, Borgani, \& Valentini}]{Fumagalli:2025twg}
Fumagalli, A., Castro, T., Borgani, S., \& Valentini, M. 2025, A\&A, 697, A140

\bibitem[{Fumagalli {et~al.}(2021)}]{Euclid:2021api}
Fumagalli, A. {et~al.} 2021, A\&A, 652, A21

\bibitem[{Hamaus {et~al.}(2016)Hamaus, Pisani, Sutter, Lavaux, Escoffier, Wandelt, \& Weller}]{Hamaus:2016wka}
Hamaus, N., Pisani, A., Sutter, P.~M., {et~al.} 2016, PRL, 117, 091302

\bibitem[{Hamaus {et~al.}(2014)Hamaus, Sutter, \& Wandelt}]{Hamaus:2014fma}
Hamaus, N., Sutter, P.~M., \& Wandelt, B.~D. 2014, PRL, 112, 251302

\bibitem[{Hamaus {et~al.}(2022)}]{Euclid:2021xmh}
Hamaus, N. {et~al.} 2022, A\&A, 658, A20

\bibitem[{Ilic {et~al.}(2013)Ilic, Langer, \& Douspis}]{Ilic:2013cn}
Ilic, S., Langer, M., \& Douspis, M. 2013, A\&A, 556, A51

\bibitem[{Jennings {et~al.}(2013)Jennings, Li, \& Hu}]{Jennings:2013nsa}
Jennings, E., Li, Y., \& Hu, W. 2013, MNRAS, 434, 2167

\bibitem[{{Kov{\'a}cs}(2018)}]{Kovacs:2018}
{Kov{\'a}cs}, A. 2018, \mnras, 475, 1777

\bibitem[{Lee \& Park(2009)}]{Lee:2007kq}
Lee, J. \& Park, D. 2009, ApJL, 696, L10

\bibitem[{{Lehman} {et~al.}(2025){Lehman}, {Schuster}, {Lucie-Smith}, {Hamaus}, {Davies}, \& {Dolag}}]{Lehman:2025}
{Lehman}, K., {Schuster}, N., {Lucie-Smith}, L., {et~al.} 2025, arXiv e-prints, arXiv:2502.05262

\bibitem[{{Li}(2011)}]{Li2011}
{Li}, B. 2011, \mnras, 411, 2615

\bibitem[{Monaco(1995)}]{Monaco:1994ed}
Monaco, P. 1995, ApJ, 447, 23

\bibitem[{Monaco(1997)}]{Monaco:1997cq}
Monaco, P. 1997, MNRAS, 287, 753

\bibitem[{Monaco {et~al.}(2013)Monaco, Sefusatti, Borgani, Crocce, Fosalba, Sheth, \& Theuns}]{Monaco:2013qta}
Monaco, P., Sefusatti, E., Borgani, S., {et~al.} 2013, MNRAS, 433, 2389

\bibitem[{Monaco {et~al.}(2002)Monaco, Theuns, \& Taffoni}]{Monaco:2001jg}
Monaco, P., Theuns, T., \& Taffoni, G. 2002, MNRAS, 331, 587

\bibitem[{Monaco {et~al.}(2025)}]{Euclid:2025lfa}
Monaco, P. {et~al.} 2025, A\&A, in press [\eprint[arXiv]{2507.12116}]

\bibitem[{Moresco {et~al.}(2022)}]{Moresco:2022phi}
Moresco, M. {et~al.} 2022, LRR, 25, 6

\bibitem[{Munari {et~al.}(2017)Munari, Monaco, Sefusatti, Castorina, Mohammad, Anselmi, \& Borgani}]{Munari:2016aut}
Munari, E., Monaco, P., Sefusatti, E., {et~al.} 2017, MNRAS, 465, 4658

\bibitem[{Neyrinck(2008)}]{Neyrinck:2007gy}
Neyrinck, M.~C. 2008, MNRAS, 386, 2101

\bibitem[{Park \& Lee(2007)}]{Park:2006wu}
Park, D. \& Lee, J. 2007, PRL, 98, 081301

\bibitem[{Perico {et~al.}(2019)Perico, Voivodic, Lima, \& Mota}]{Perico:2019obq}
Perico, E. L.~D., Voivodic, R., Lima, M., \& Mota, D.~F. 2019, A\&A, 632, A52

\bibitem[{Pisani {et~al.}(2015)Pisani, Sutter, Hamaus, Alizadeh, Biswas, Wandelt, \& Hirata}]{Pisani:2015jha}
Pisani, A., Sutter, P.~M., Hamaus, N., {et~al.} 2015, PRD, 92, 083531

\bibitem[{Pisani {et~al.}(2019)}]{Pisani:2019cvo}
Pisani, A. {et~al.} 2019 [\eprint[arXiv]{1903.05161}]

\bibitem[{Platen {et~al.}(2007)Platen, van~de Weygaert, \& Jones}]{Platen:2007qk}
Platen, E., van~de Weygaert, R., \& Jones, B. J.~T. 2007, MNRAS, 380, 551

\bibitem[{Press \& Schechter(1974)}]{Press:1973iz}
Press, W.~H. \& Schechter, P. 1974, ApJ, 187, 425

\bibitem[{Radinovic {et~al.}(2023)}]{Euclid:2023eom}
Radinovic, S. {et~al.} 2023, A\&A, 677, A78

\bibitem[{Sachs \& Wolfe(1967)}]{Sachs:1967er}
Sachs, R.~K. \& Wolfe, A.~M. 1967, ApJ, 147, 73

\bibitem[{{Schuster} {et~al.}(2023){Schuster}, {Hamaus}, {Dolag}, \& {Weller}}]{Schuster:2022ogh}
{Schuster}, N., {Hamaus}, N., {Dolag}, K., \& {Weller}, J. 2023, JCAP, 2023, 031

\bibitem[{{Schuster} {et~al.}(2024){Schuster}, {Hamaus}, {Dolag}, \& {Weller}}]{Schuster:2024}
{Schuster}, N., {Hamaus}, N., {Dolag}, K., \& {Weller}, J. 2024, JCAP, 2024, 065

\bibitem[{{Schuster} {et~al.}(2019){Schuster}, {Hamaus}, {Pisani}, {Carbone}, {Kreisch}, {Pollina}, \& {Weller}}]{Schuster:2019hyl}
{Schuster}, N., {Hamaus}, N., {Pisani}, A., {et~al.} 2019, JCAP, 2019, 055

\bibitem[{{Schuster} {et~al.}(2025){Schuster}, {Hamaus}, {Pisani}, {Dolag}, \& {Weller}}]{Schuster:2025}
{Schuster}, N., {Hamaus}, N., {Pisani}, A., {Dolag}, K., \& {Weller}, J. 2025, arXiv e-prints, arXiv:2509.07092

\bibitem[{Shandarin \& Zeldovich(1989)}]{Shandarin:1989sr}
Shandarin, S.~F. \& Zeldovich, Y.~B. 1989, RMP, 61, 185

\bibitem[{Sheth \& Tormen(1999)}]{Sheth:1999mn}
Sheth, R.~K. \& Tormen, G. 1999, MNRAS, 308, 119

\bibitem[{Sheth \& van~de Weygaert(2004)}]{Sheth:2003py}
Sheth, R.~K. \& van~de Weygaert, R. 2004, MNRAS, 350, 517

\bibitem[{Springel(2005)}]{Springel:2005mi}
Springel, V. 2005, MNRAS, 364, 1105

\bibitem[{Springel {et~al.}(2001)Springel, White, Tormen, \& Kauffmann}]{Springel:2000qu}
Springel, V., White, S. D.~M., Tormen, G., \& Kauffmann, G. 2001, MNRAS, 328, 726

\bibitem[{Sutter {et~al.}(2015)Sutter, Lavaux, Hamaus, Pisani, Wandelt, Warren, Villaescusa-Navarro, Zivick, Mao, \& Thompson}]{Sutter:2014haa}
Sutter, P.~M., Lavaux, G., Hamaus, N., {et~al.} 2015, A\&C, 9, 1

\bibitem[{Sutter {et~al.}(2014)Sutter, Pisani, Wandelt, \& Weinberg}]{Sutter:2014oca}
Sutter, P.~M., Pisani, A., Wandelt, B.~D., \& Weinberg, D.~H. 2014, MNRAS, 443, 2983

\bibitem[{Warren {et~al.}(2006)Warren, Abazajian, Holz, \& Teodoro}]{Warren:2005ey}
Warren, M.~S., Abazajian, K., Holz, D.~E., \& Teodoro, L. 2006, ApJ, 646, 881

\bibitem[{Watson {et~al.}(2013)Watson, Iliev, D'Aloisio, Knebe, Shapiro, \& Yepes}]{Watson:2012mt}
Watson, W.~A., Iliev, I.~T., D'Aloisio, A., {et~al.} 2013, MNRAS, 433, 1230

\bibitem[{Xu(1995)}]{Xu:1994fk}
Xu, G.-H. 1995, ApJS, 98, 355

\end{thebibliography}

\begin{appendix}
\section{Resolution comparison}\label{Resolution_comparison}
In this appendix, we present a resolution comparison, i.e., \PINOCCHIO\ vs. \PINOCCHIO\ and \og\ vs. \og, for two cosmic void summary statistics analyzed in the main text, where resolution effects appear to be more significant.

\begin{figure}[t]
    \centering
    \includegraphics[width=1.0\textwidth]{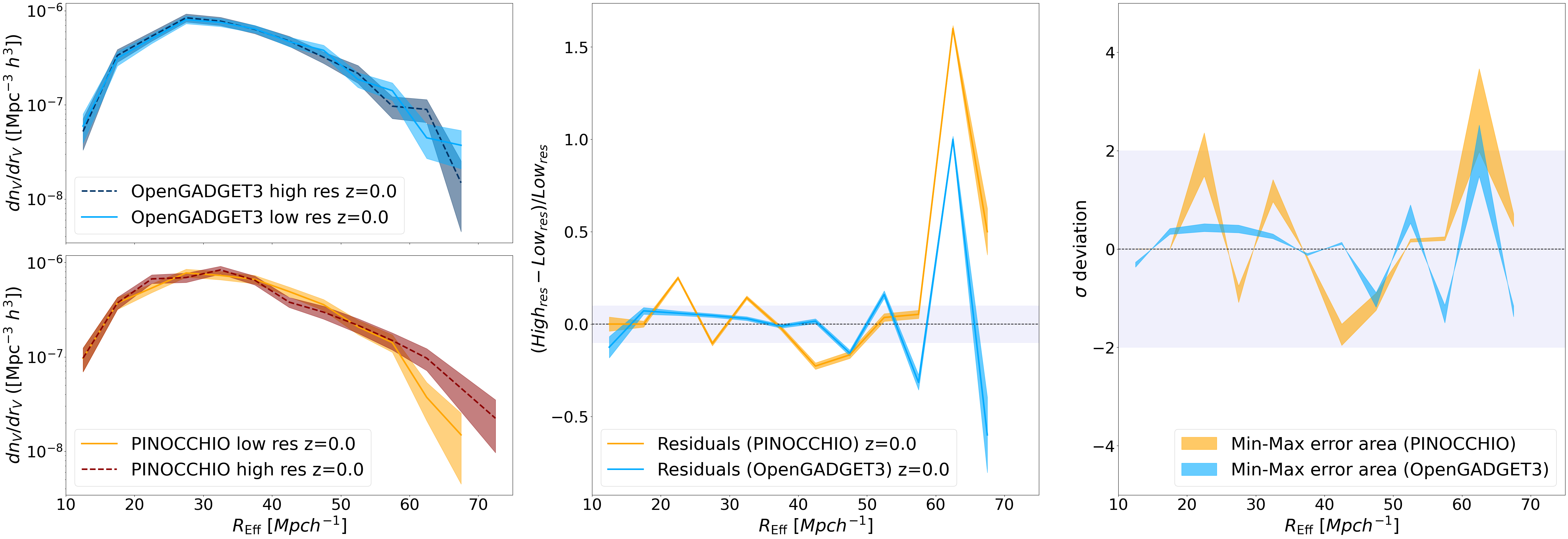}
    
    \caption{Same as Figure \ref{VSF_comparison} for VSFs across different resolutions.}
    \label{VSF_resolution_comparison}
\end{figure}

In particular, Figure~\ref{VSF_resolution_comparison} shows the same measurement as in Figure~\ref{VSF_comparison}, but now comparing different resolutions within each code. The variation in the VSF with resolution is more pronounced in \PINOCCHIO\ than in \og, which remains largely stable. This suggests a stronger resolution dependence in \PINOCCHIO.

\begin{figure}[t]
    \centering
    \includegraphics[width=1.0\textwidth]{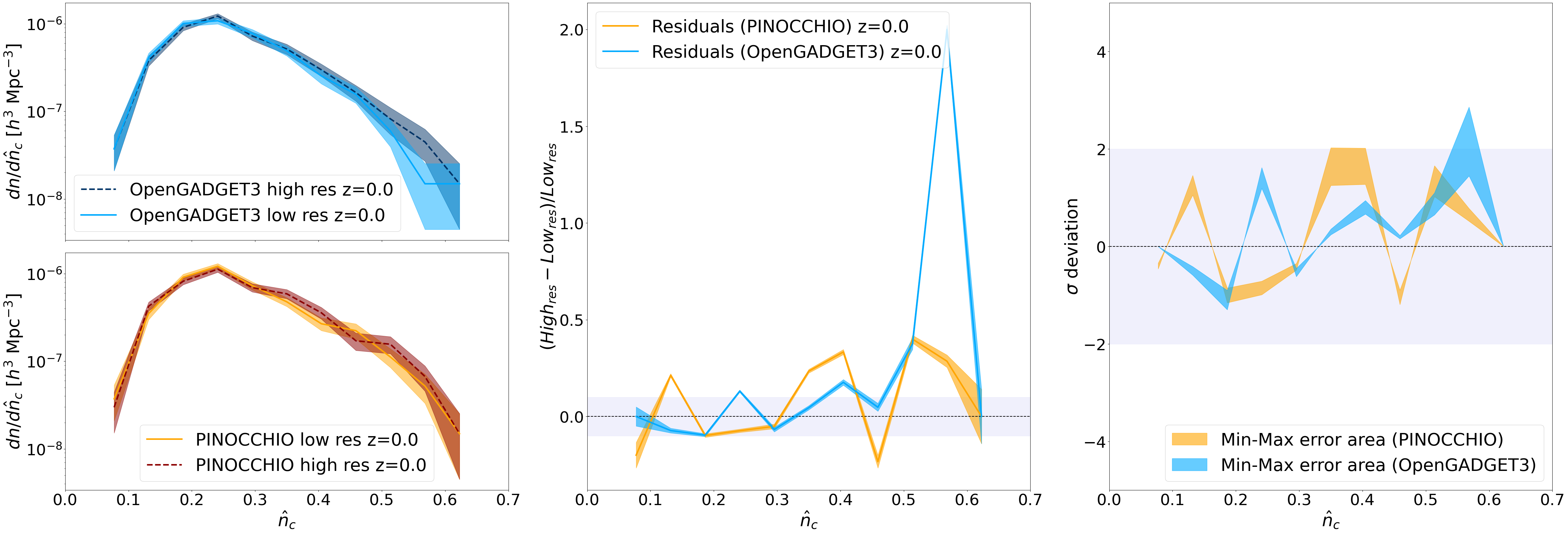}
    
    \caption{Same as Figure \ref{CoreDens_comparison} for CDFs across different resolutions.}
    \label{CDF_resolution_comparison}
\end{figure}

Similarly, Figure~\ref{CDF_resolution_comparison} shows the same measurement as in Figure~\ref{CoreDens_comparison}, again comparing resolutions within each code. The resolution dependence of the discrepancy supports the interpretation above: agreement between the two codes improves at higher resolution, where the \frag\ algorithm is expected to operate more effectively.

\clearpage
\section{Large box simulation}\label{Large_Box}

Since running new pairs of simulations is computationally expensive due to the high cost of \og\ runs, we make use of an already available set of larger-box simulations to test the robustness of our analysis with an increased sample of voids. As in the main analysis, the ICs for these simulations were generated with \PINOCCHIO\ using 3LPT at $z = 50$ . 

The simulations have a box size of $3870h^{-1}\mathrm{Mpc}$ and a particle resolution of $2048^3$. The overall analysis presented here follows exactly the same procedure described in the main text, with the only differences being the higher halo mass cut of $10^{14}M_\odot$, required due to the lower mass resolution compared to our smaller boxes, and the availability of only three redshift snapshots ($z=0.0, 0.5, 1.0$).

\begin{figure}[t]
    \centering
    \includegraphics[width=1.0\columnwidth]{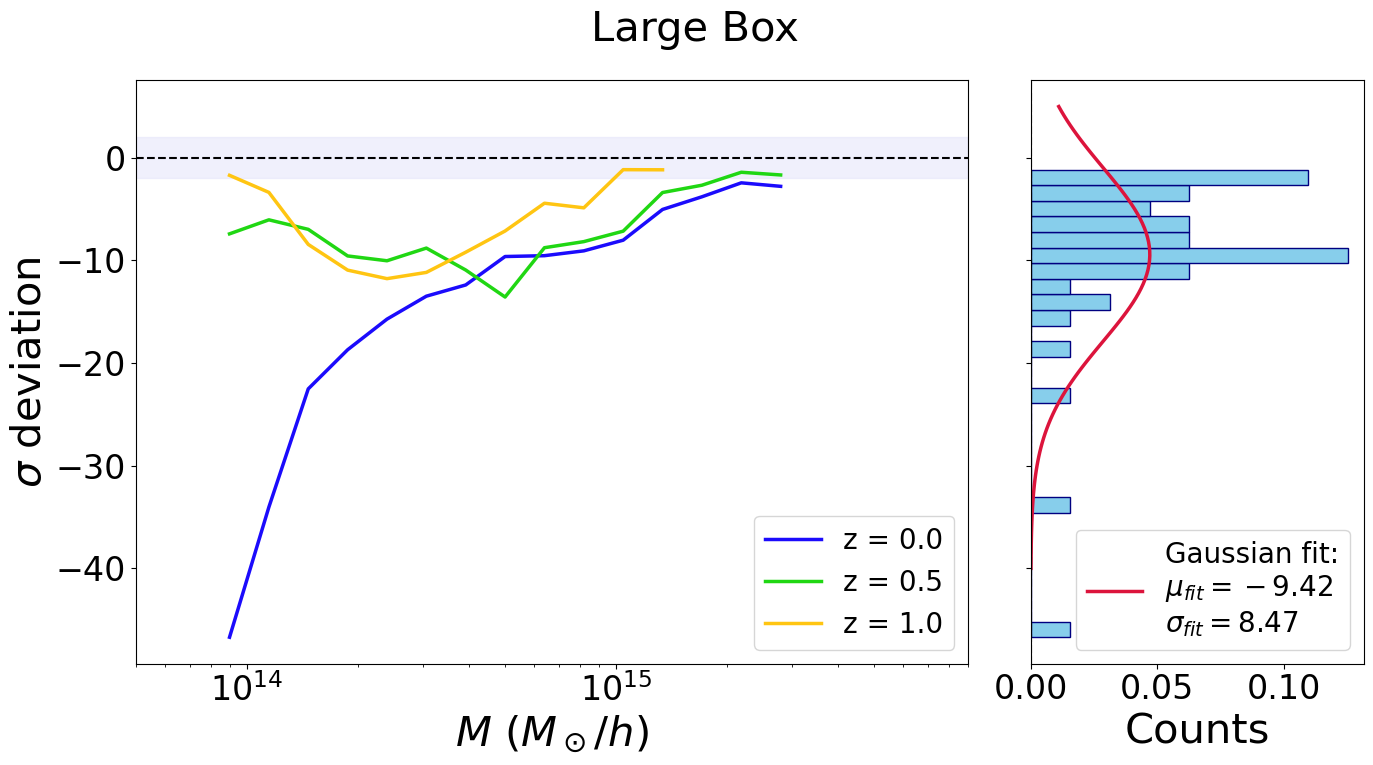}
    
    \caption{Same as Figure \ref{HMF_evolution}, but for the large box.}
    \label{HMF_evolution_large}
\end{figure}

\begin{figure*}[t]
    \centering
    \includegraphics[width=1.0\textwidth]{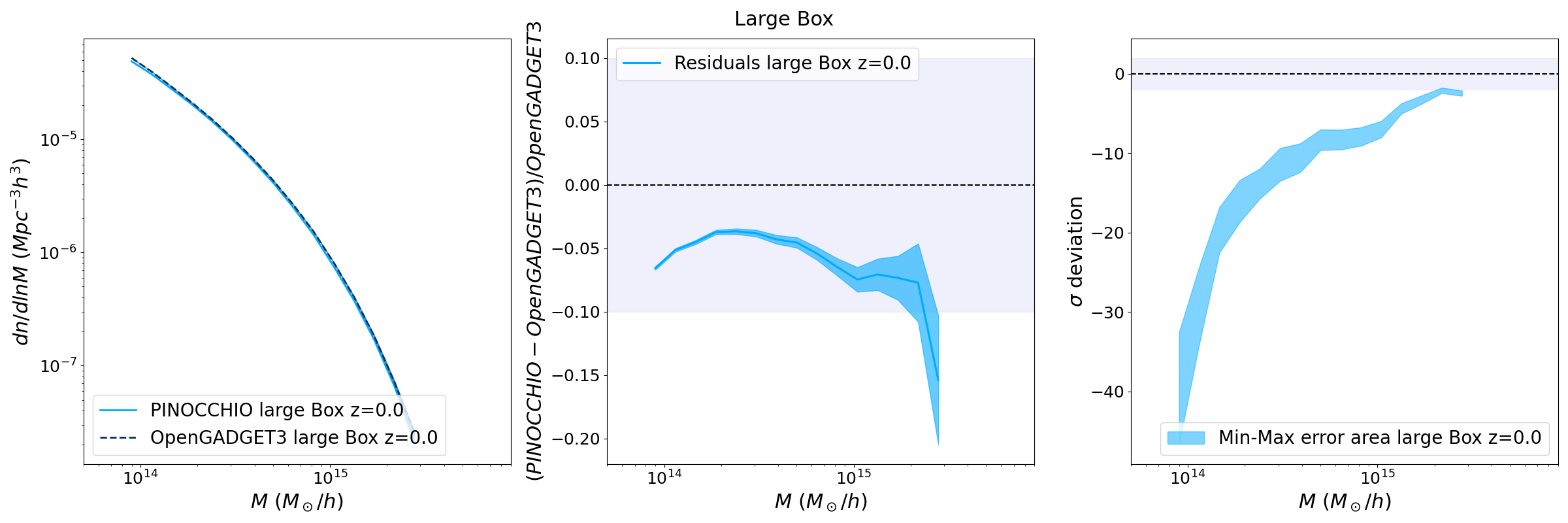}
    
    \caption{Same as Figure \ref{HMF_comparison}, but for the large box.}
    \label{HMF_large}
\end{figure*}

Figures~\ref{HMF_large} and~\ref{HMF_evolution_large} show, respectively, the same measurements as Figures~\ref{HMF_comparison} and~\ref{HMF_evolution}, but for the larger box. These results confirm the clear mass dependence of the deviations already discussed in Section~\ref{HMF_comparison_section}, with systematic discrepancies at the low-mass end where halo identification in \PINOCCHIO\ is less reliable. Nonetheless, as in the main analysis, the agreement between the two methods remains within 10\%, suggesting that we are limited by systematic effects in this regime.

Despite the systematic deviations in the HMF, the void summary statistics remain consistent with the main analysis. The number of voids identified for each redshift $z$ is reported in Table~\ref{tab:void_counts_large}, showing comparable counts between the two codes, in agreement with the findings of the smaller-box runs.

In addition, we have measured the VSF, VEFs, CDFs, and RDPs, together with the relative differences between \PINOCCHIO\ and \og, normalized by the Jackknife errors from the \og\ estimates at the three available redshifts.

\begin{figure}[h!]
    \centering
    \includegraphics[width=1.0\columnwidth]{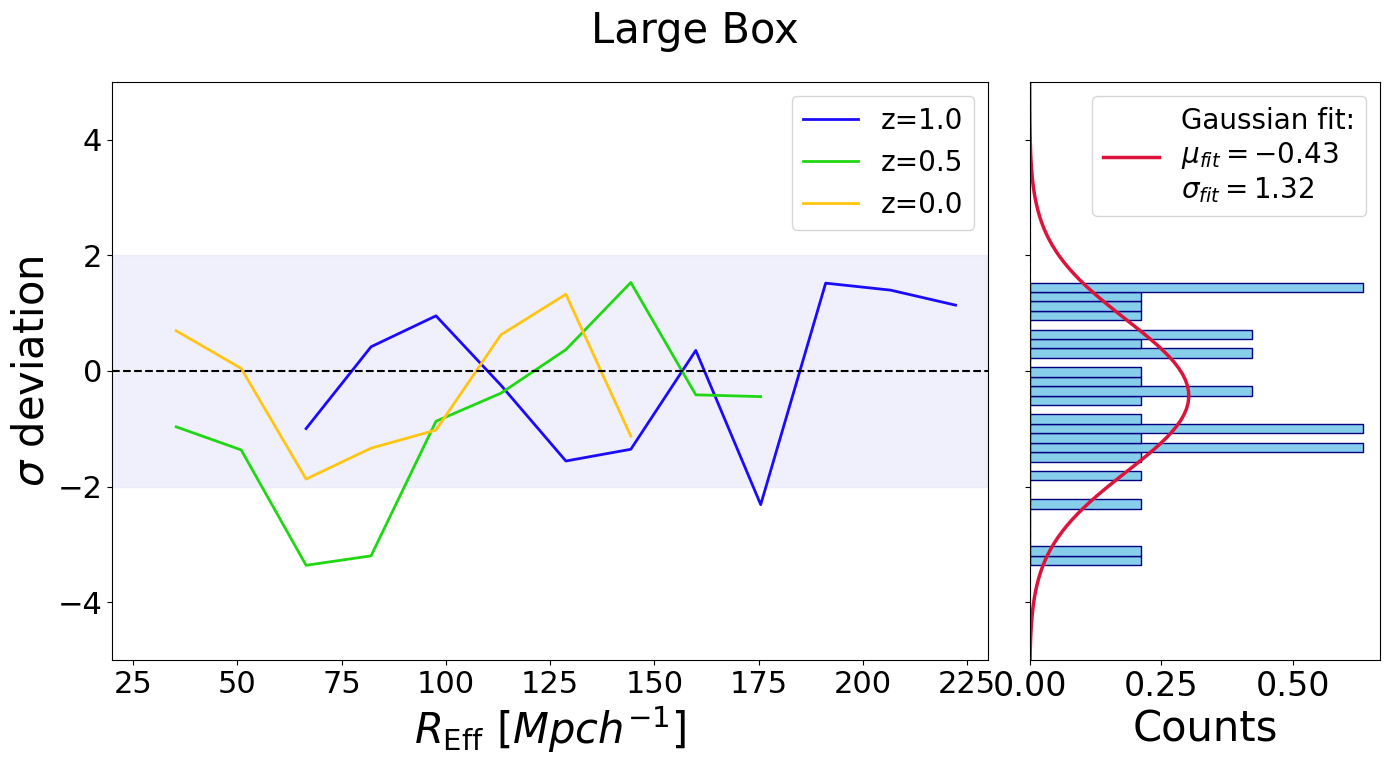}
    
    \caption{Same as Figure \ref{VSF_evolution}, but for the large box.}
    \label{VSF_evolution_large}
\end{figure}

\begin{figure}[h!]
    \centering
    \includegraphics[width=1.0\columnwidth]{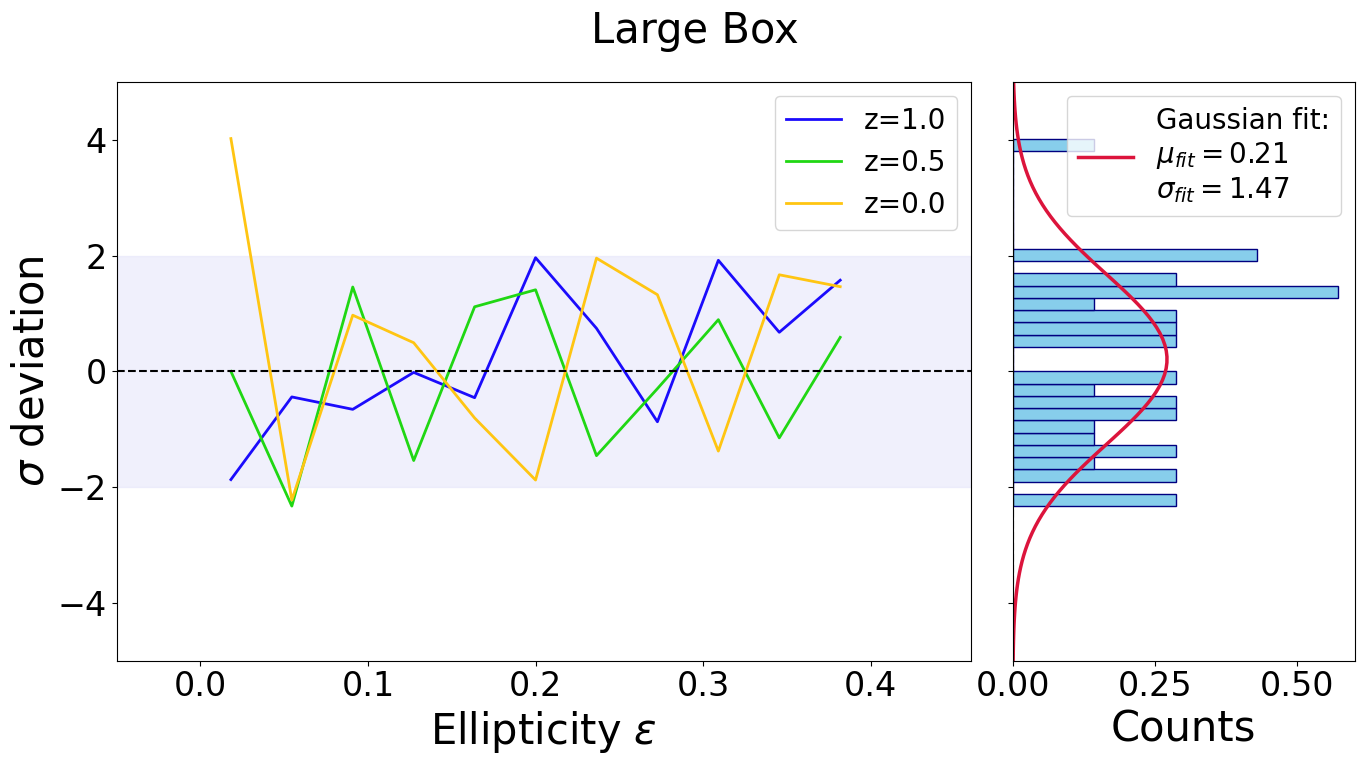}
    
    \caption{Same as Figure \ref{Ellipticity_evolution}, but for the large box.}
    \label{Ellipticity_evolution_large}
\end{figure}

\begin{figure*}[t]
    \centering
    \includegraphics[width=1.0\textwidth]{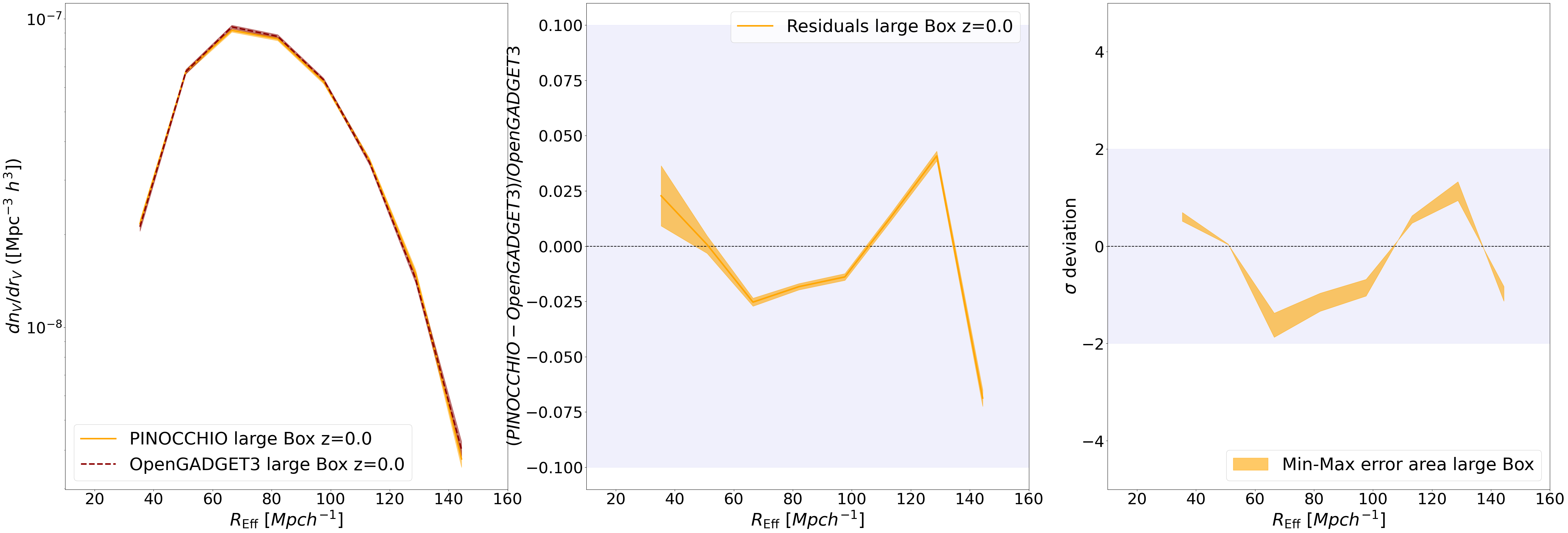}
    
    \caption{Same as Figure \ref{VSF_comparison}, but for the large box.}
    \label{VSF_large}
\end{figure*}

\begin{figure*}[t]
    \centering
    \includegraphics[width=1.0\textwidth]{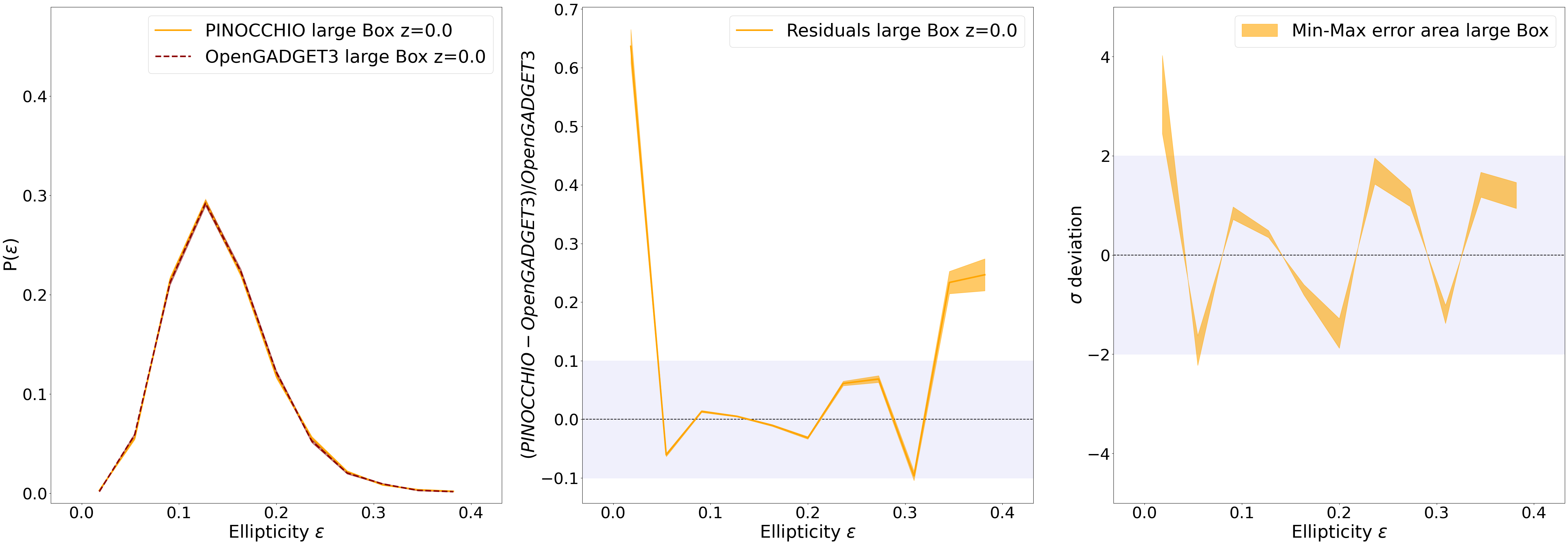}
    
    \caption{Same as Figure \ref{Ellipticity_comparison}, but for the large box.}
    \label{Ellipticity_large}
\end{figure*}

\begin{figure*}[t]
    \centering
    \includegraphics[width=1.0\textwidth]{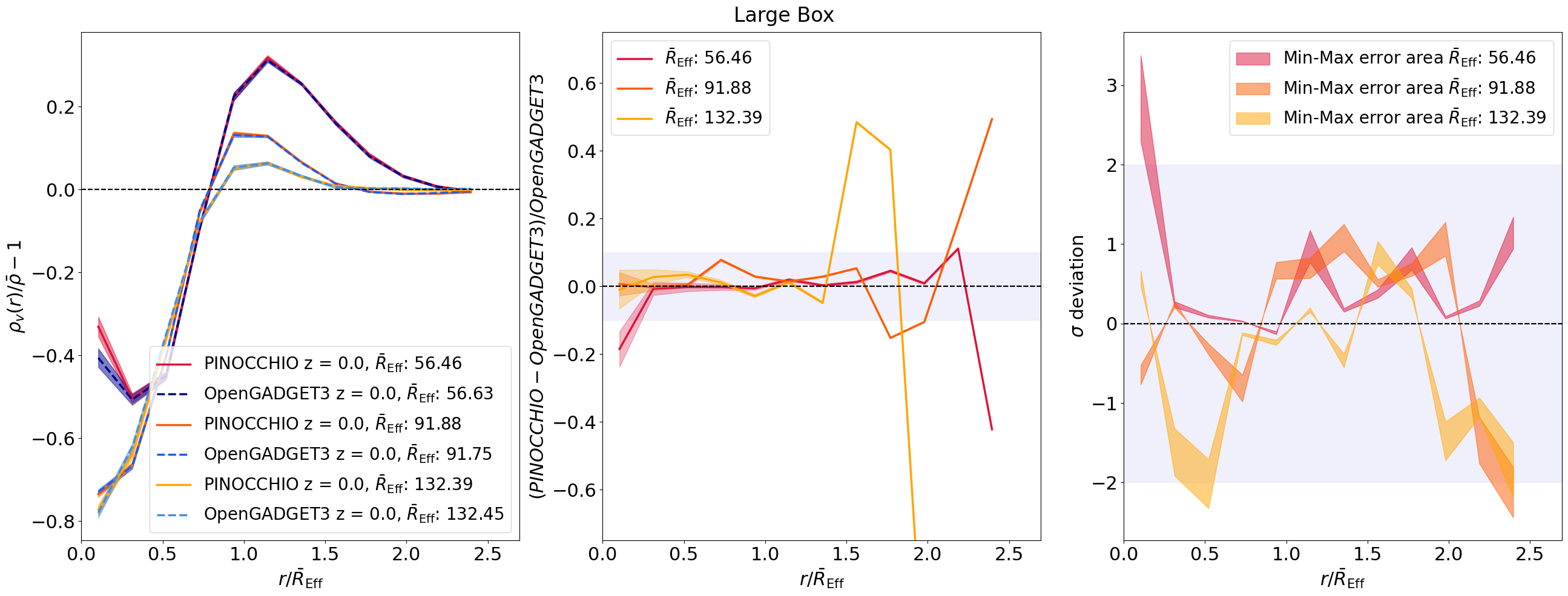}
    
    \caption{Same as Figure \ref{Profile_comparison}, but for the large box.}
    \label{Profile_large}
\end{figure*}

\begin{figure*}[t]
    \centering
    \includegraphics[width=1.0\textwidth]{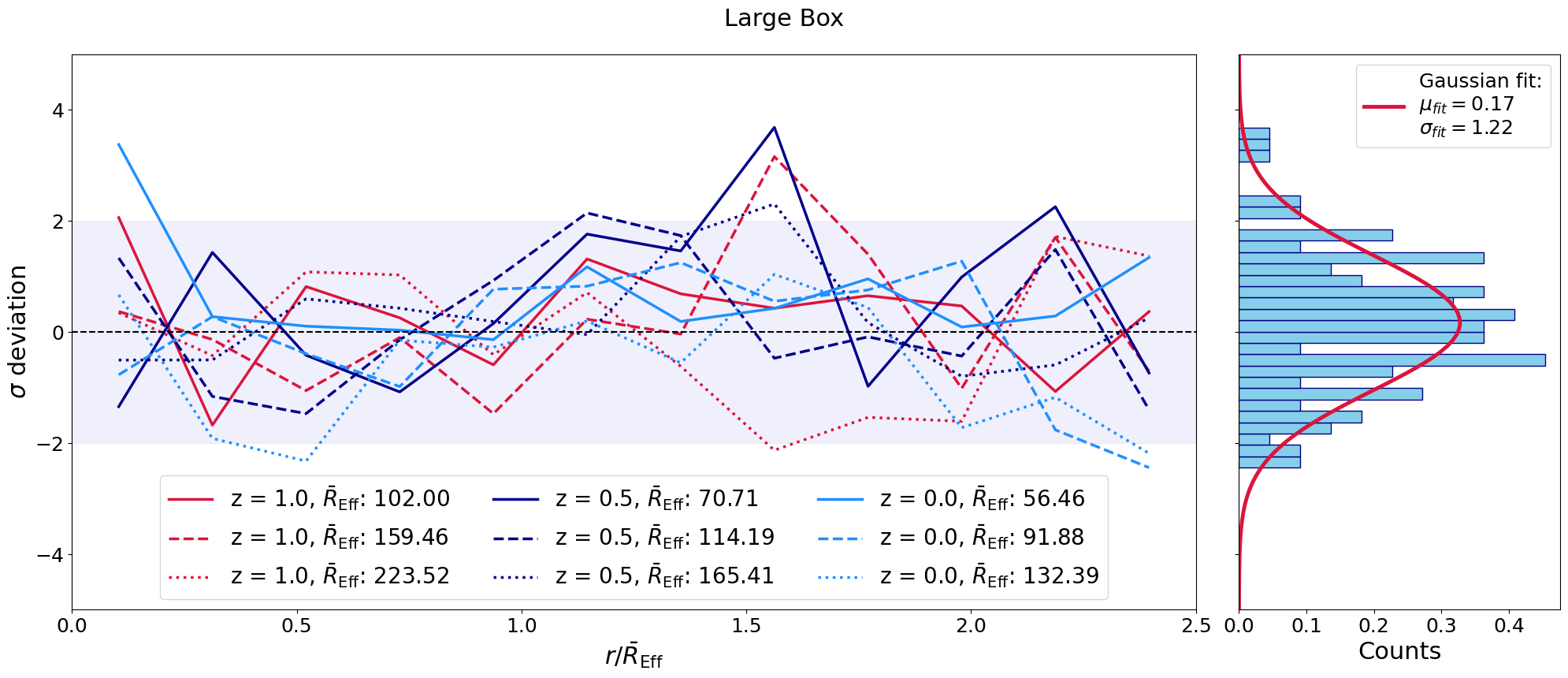}
    
    \caption{Same as Figure \ref{Profile_evolution}, but for the large box.}
    \label{Profile_evolution_large}
\end{figure*}

The results for the VSF are shown in Figures~\ref{VSF_large} and~\ref{VSF_evolution_large}, and confirm the same qualitative trends discussed in the main analysis. The same holds for the VEFs and RDPs, shown respectively in Figures~~\ref{Ellipticity_large},~\ref{Ellipticity_evolution_large},~\ref{Profile_large} and~\ref{Profile_evolution_large}.

\begin{figure}[h!]
    \centering
    \includegraphics[width=1.0\columnwidth]{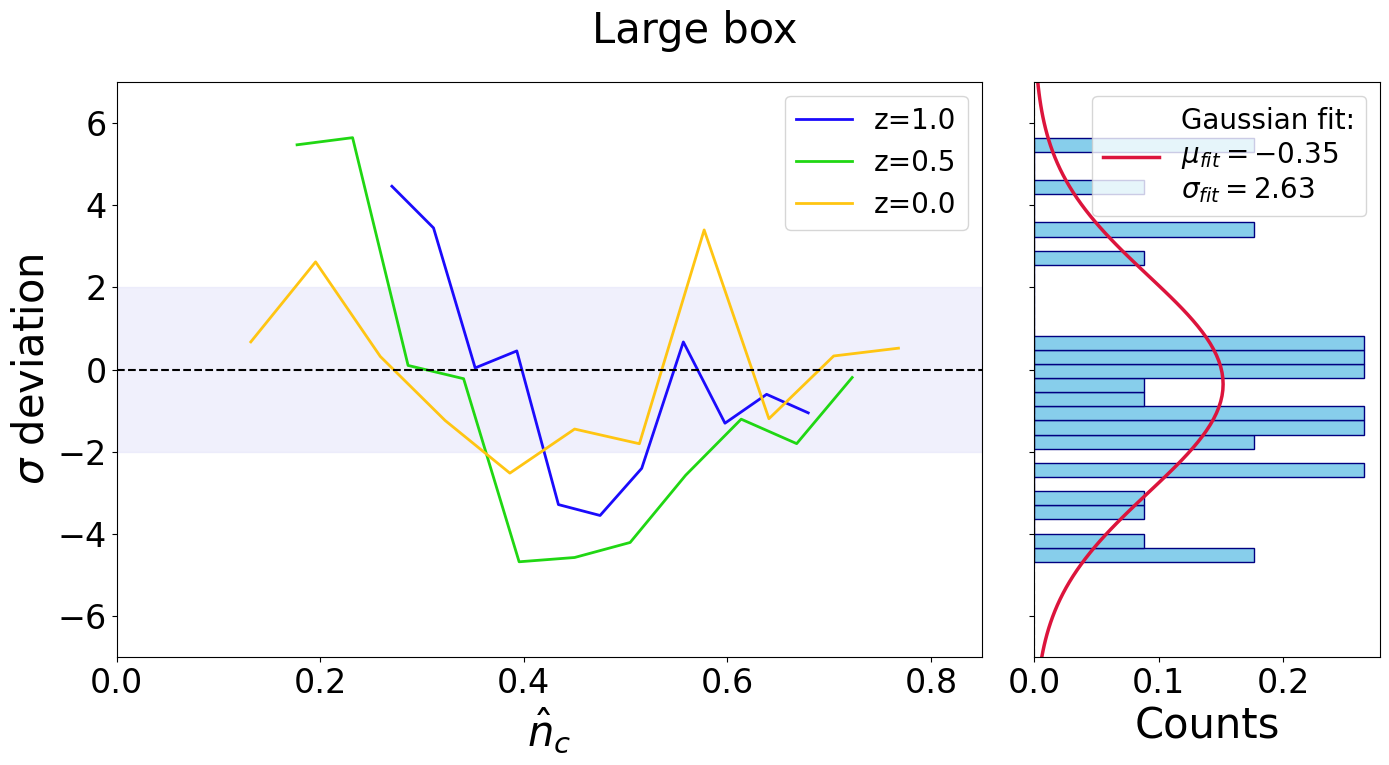}
    
    \caption{Same as Figure \ref{CoreDens_evolution}, but for the large box.}
    \label{Core_evolution_large}
\end{figure}

\begin{figure*}[t]
    \centering
    \includegraphics[width=1.0\textwidth]{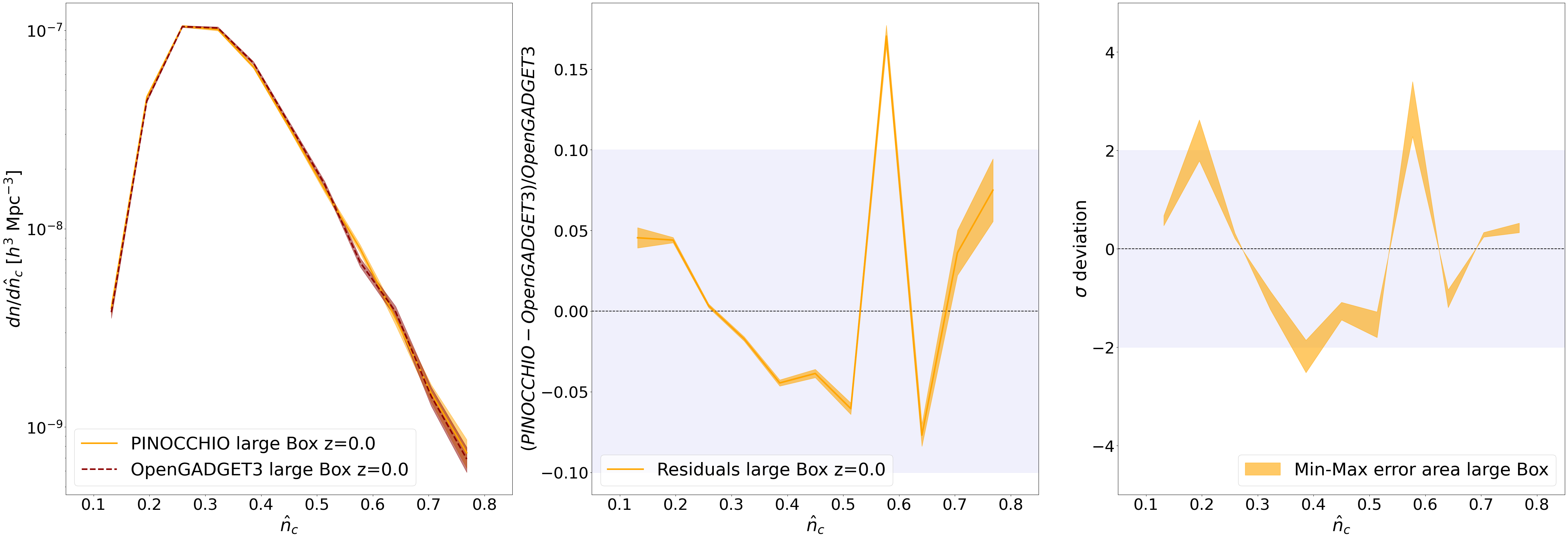}
    
    \caption{Same as Figure \ref{CoreDens_comparison}, but for the large box.}
    \label{CoreDens_large}
\end{figure*}

The CDFs, presented in Figures~\ref{CoreDens_large} and~\ref{Core_evolution_large}, represent the only statistic where the agreement worsens, particularly at redshifts $z>0$. However, this behavior is consistent with the discussion in the main text and in Appendix~\ref{Resolution_comparison}: the agreement in the CDFs improves with increasing resolution. Since the large-box runs have lower mass resolution than even our “low-resolution” simulations, the larger discrepancies found here are expected.

Overall, these results confirm the robustness of the analysis presented in the main text, while also indicating that increasing the statistical sample of analyzed voids can significantly improve the precision of the measured summary statistics.
\begin{table}[t]
\centering
\caption{Number of voids identified by \vide}
\label{tab:void_counts_large}
\resizebox{0.78\columnwidth}{!}{%
\begin{tabular}{c|c|c}
\toprule
$z$ & {\PINOCCHIO} & {\og} \\
             & Large Box & Large Box \\
\midrule
0.0 & 22302 & 22506 \\
0.5 & 10794 & 11131 \\
1.0 & 3564 & 3617 \\
\bottomrule
\end{tabular}%
}
\tablefoot{The first column lists the redshifts $z$, while the second and third columns shows the number of voids identified for both simulation resolutions and codes.}
\end{table}

\end{appendix}
\end{document}